\documentclass[a4paper,12pt]{article}
\pdfoutput=1 

\usepackage[T1]{fontenc} 
\usepackage{amsmath,amssymb,latexsym}
\usepackage[english]{babel}
\usepackage{graphicx}
\usepackage[T2A]{fontenc}
\usepackage[utf8]{inputenc}
\usepackage{overpic}
\usepackage[unicode,colorlinks,linkcolor=blue,citecolor=red]{hyperref} 

\def\nqq{\hspace*{-2em}}

\def\lal{&&\nqq {}}
\def\eq{Eq.\,}
\def\eqs{Eqs.\,}
\def\beq{\begin{equation}}
\def\eeq{\end{equation}}
\def\bear{\begin{eqnarray}}
\def\bearr{\begin{eqnarray} \lal}
\def\ear{\end{eqnarray}}
\def\earn{\nonumber \end{eqnarray}}
\def\nn{\nonumber\\ {}}

\def\dst{\displaystyle}
\def\tst{\textstyle}
\def\fracd#1#2{{\dst\frac{#1}{#2}}}
\def\fract#1#2{{\tst\frac{#1}{#2}}}
\def\Half{{\fracd{1}{2}}}
\def\half{{\fract{1}{2}}}

\def\e{{\,\rm e}}
\def\d{\partial}

\def\const{{\rm const}}

\def\eqn#1{\eq\eqref{#1}}
\def\rf{\eqref}
\def\mn{_{\mu\nu}}

\def\kappa{\varkappa}

\def\M{{\mathbb M}}

\title{\bf Inhomogeneous compact extra dimensions and de Sitter cosmology}

\author{{Kirill A. Bronnikov}$^{a,b,c}$\thanks{e-mail: kb20@yandex.ru} \and 
{Arkady~A.~Popov}$^{d}$\thanks{e-mail: arkady\_popov@mail.ru} \and
{Sergey~G.~Rubin}$^{c,d}\thanks{e-mail: sergeirubin@list.ru}$}

\date{\em \small 
$^a$ VNIIMS, 119361 Ozyornaya ul. 46, Moscow, Russia\\
$^b$ Inst. of Gravitation and Cosmology, RUDN University,  117198 ul. Miklukho-Maklaya 6, Moscow, Russia \\
$^c$ National Research Nuclear University MEPhI (Moscow Engineering Physics Institute), 115409, Kashirskoe shosse 31, Moscow, Russia \\
$^d$ N.~I.~Lobachevsky Institute of Mathematics and Mechanics, Kazan  Federal  University, 420008, Kremlevskaya  street  18,  Kazan,  Russia}

\begin{document}
\maketitle
\begin{abstract}In the framework of multidimensional $f(R)$ gravity,
  we study the metrics of compact extra dimensions assuming that our 4D space has the
  de Sitter metric. Manifolds described by such metrics could be formed at the inflationary
  and even higher energy scales. It is shown that in the presence of a scalar field, varying
  in the extra factor space $\M_2$, it is possible to obtain a variety of inhomogeneous metrics
  in $\M_2$. Each of these metrics leads to a certain value of the 4D cosmological
  constant $\Lambda_4$, and in particular, it is possible to obtain $\Lambda_4 =0$, as
  is confirmed by numerically obtained solutions. A nontrivial scalar field distribution in the
  extra dimensions is an important feature of this family of models.
  \end{abstract}

\section{Introduction}

  The idea of extra dimensions is widely used for explanation of a variety of phenomena, such as
  the physics beyond the Standard Model, cosmological scenarios including inflationary models
  and the origin of the dark components of the Universe, etc.
  \cite{Abbott:1984ba,Chaichian:2000az,Randall:1999vf,Brown:2013fba}.
  Sometimes extra dimensions are endowed with scalar fields and antisymmetric form fields
  to stabilize their metric. There are models where the Casimir effect is taken into account
  \cite{Fischbach:2009ud, 2018GrCo...24..154B}. Thus inclusion of extra dimensions
  is a promising background for the physics below $\sim 10$ TeV.

  At the same time, it is usually assumed that the Universe has been nucleated due to
  quantum processes at very high energies
  \cite{1994CQGra..11.2483V,Bousso:1998ed}.
  The metric of our Universe and the fields inside the horizon experience strong quantum
  fluctuations that could affect their dynamics and their final states at low energies
  \cite{Brandenberger:2006vv,Tegmark:2005dy}, including the shape of extra dimensions.

  In this paper, we study the possible influence of a matter field on the metric of extra dimensions.
  Previous results concerning multidimensional gravity with 4D Minkowski factor space are
  published in \cite{Gani:2014lka,Rubin:2016ude,Rubin:2015pqa,Lyakhova:2018zsr,Nikulin2019},
  where the importance of inhomogeneous extra dimensions was discussed. Inparticular, the scalar field localization on deformed extra space.
  Here we extend the same idea to a more general case, the 4D de Sitter metric with an
  arbitrary Hubble parameter. Such a metric is approximately realized at the inflationary stage
  and can be valid up to Planck energies.

The models under consideration contain two extra dimensions that form a compact surface
  of rotation, which in the general case possesses a conical singularity at a particular point
  (``the second center''). The scalar field is to a large extent concentrated in a neighborhood of
  this point, showing a behavior similar to what is observed in many brane-world models. However,
  unlike such models, we assume that the size of extra dimensions is small enough to be invisible
  in modern accelerator experiments, i.e., we actually adhere to the Kaluza-Klein concept of extra
  dimensions. A narrower class of models are completely regular, however, it should be noted
  that their basic physical properties, including their 4D appearance, are almost indistinguishable
  from those of models with conical singularities.

  Our study is based on multidimensional $f(R)$ gravity. The interest in $f(R)$ theories is
  motivated by inflationary scenarios starting with the work of Starobinsky \cite{Starobinsky:1980te}.
  Having been developed 40 yeas ago on the basis of 4D $R^2$ gravity, this model remains
  most promising up to now.  Any combination of quantities invariant under the general coordinate
  transformations may be used in the theory if one keeps in mind two issues. Firstly, a theory must
  restore the Einstein-Hilbert action at low energies. Secondly, any gravitational action including
  the Einstein-Hilbert one is nonrenormalizable and should be considered as an effective theory.
	
  The simplest extension of general relativity is the one containing a function of the Ricci scalar
  $f(R)$. Some viable $f(R)$ models in 4D space that satisfy the observational constraints have
  been proposed in \cite{DeFelice:2010aj,2014JCAP...01..008B,Nojiri_2017}. Such modified gravity
  can be responsible for dark energy \cite{2007PhLB..651..224N}.

  Stabilization of the extra space as a pure gravitational effect in $f(R)$ and more general
  multidimensional theories with maximally symmetric extra spaces has been studied in
  \cite{Bronnikov:2005iz,BronRub,2003Ap&SS.283..679G}, as well as their ability
  to describe both early and late inflationary expansion
  \cite{Bronnikov:2008cr,Bronnikov:2009zza,Bronnikov:2009ai}.
  In \cite{Rubin:2015pqa}, it was shown that an $f(R)$ model with inhomogeneous extra space is
  compatible with 4D Minkowski or very weakly curved space-times.

  The structure of this paper is as follows. In Section 2 we choose the metric and dimensionality
  of our manifold, the Lagrangian containing gravity with higher derivatives and a scalar field
  and derive the set of classical equations. In Section 3 we discuss the boundary conditions
  that are necessary in order to obtain a set of solutions and present a number of numerical
  solutions obtained under these conditions. Section 4 is devoted to a stability study for the obtained solutions. In Section 5 we discuss the 4D properties of these solutions.
  Conclusion are made in Section 6.

\section{Basic equations}

  We will consider 6D metrics of the general form
\beq           \label{ds0}
	ds^2 = g\mn dx^\mu dx^\nu  - \e^{2\alpha(u)} du^2 - \e^{2\beta(u)} d\varphi^2,
\eeq		
  where $u$ and  $\varphi$ are extra spatial coordinates, and $\varphi \in [0, 2\pi)$
  is assumed to be a polar angle, while the 4D metric tensor $g\mn$ may depend on
  both 4D coordinates (making it possible to consider, for example, cosmological
  or static models) and the ``radial'' fifth coordinate $u$.
  The extra factor space $\M_2$ parametrized by $(u, \varphi)$ is thus a surface of rotation,
  which can be compact if
  the circular radius $r(u)\equiv \e^{\beta(u)}$ tends to zero at two boundary values of $u$. The 6-dimensional metric is chosen as the simplest nontrivial metric suitable for our purposes.

  In such space-times, we consider a theory with the action
\bear                           \label{S}
	    S &=&  \int d^6 x \sqrt{|g_6|} \bigg[ \frac{m_D^4}{2} f(R)
	     +  \Half g^{AB} \phi_{,A}\phi_{,B} - V(\phi) \bigg],  \nn
       && A, B = 0, \ldots, 5.
\ear  	  	
  where $g_6 = \det (g_{AB})$,  $f(R)$ and $V(\phi)$ are some functions (to be chosen
  later) of the 6D scalar  curvature $R$ and the scalar field $\phi$, respectively.
  Variation of \rf{S} with respect to $\phi$ and $g^{AB}$ leads to the field equations
\beq
	    \Box \phi +  V_\phi =0, \quad {\rm where}\
	    \Box\phi = \nabla_A \nabla^A \phi, \quad  V_\phi = dV/d\phi,	\label{eq-phi}
\eeq
\bear
	   && -\Half \delta_A^B f (R) + \big[ R^B_A + \nabla_A \nabla^B - \delta^B_A \Box \big] f_R
	        = - \frac{1}{m_D^2} T^B_A,
	    \nn && \hskip3mm f_R = df/dR,	                                                                      \label{EE}
\ear		  		  		  		  		  		  		  		
  and the stress-energy tensor of the scalar field $\phi = \phi (y)$ reads
\beq
		T^B_A[\phi] = \phi_{,A}\phi^{,B} - \half \delta^B_A \phi_{,C}\phi^{,C} + \delta^B_A V.
\eeq

  Before writing the particular equations to be solved, it is helpful to present
  brief expressions for the Ricci tensor components $R^A_B$ assuming a general
  diagonal metric in arbitrary dimensions,
\beq               \label{ds1}
		ds^2 = \sum_A \eta_A \e^{2b_A(X)}  (dx^A)^2, \qquad A = 0, \ldots, D-1,
\eeq
  where $b_A(X)$ are arbitrary functions of $x^A$, and $\eta_A = \pm 1$.
  Then for the diagonal components of $R^A_B$ we have
\beq                       \label{Ric_MM}
		R^{\underline M}_M =\hskip-2mm\sum_{A\ne M}\hskip-1mm\bigg( \Box_A b_M + \Box_M b_A
		- \eta_M \e^{-2b_M} b_{A,M} \sum_{B \ne A,M}\hskip-1mmb_{B,M} \bigg),
\eeq
  where no summing is assumed over an underlined index,
\[
               b_{A,B} = \d_B b_A ,\qquad
               \Box_M f(X) =  \frac 1{\sqrt{g}} \d_M \Big(\sqrt{g} g^{\underline{M}M} \d_M f(X)\Big)
\]
  for an arbitrary function $f(X)$, and $g = |\det(g_{MN})|$. The off-diagonal components
  of the Ricci tensor are more conveniently written with lower indices, namely,
\beq                        \label{Ric_MN}
	  R_{MN}=\hskip-2mm \sum_{A \ne M,N}\hskip-2mm\Big(b_{A,MN}+b_{A,M} b_{A,N}
	  			-b_{M,N} b_{A,M}-b_{N,M} b_{A,N} \Big).	
\eeq

  In the present study, we consider a cosmological (de Sitter) metric in 4D space-time
  and the extra dimensions using the Gaussian $u$ coordinate (length along the coordinate
  axis of $x^4 =u$), so that the metric \rf{ds0} takes the form
\bear                      \label{ds2}
		ds^2 &=& dt^2 -\e^{2 H t} \delta_{i j} dx^i dx^j  -  d u^2 - r(u)^2 d\varphi^2,
		\nn &&  i, j = 1,2,3,
\ear
    where $H = \const$ is the Hubble constant.  Accordingly, in terms of \rf{ds1} we now have
\beq                  \label{b_A}
             b_0 = 0, \ \ \ b_i = H t, \ \ \ b_4 = 0, \ \ \ b_5 = \ln r(u),
\eeq
   and the expressions for $R^A_B$ are greatly simplified: the only nonzero components
   of $R^A_B$ and the scalar $R$ are (the prime stands for $d/du$)
\beq                 \label{Ric2}
		R^t_t =  R^{\underline i}_i = 3H^2, \qquad
		 R^u_u = R^\varphi_\varphi = - \frac {r''}{r},	
\eeq
\beq		        \label{R2}
		R = 12 H^2 - \frac{2 r''}{r},
\eeq		
   Assuming $\phi = \phi(u)$, \eq \rf{eq-phi}  and noncoinciding equations from \rf{EE} may be
   written as
\bear              \label{phi''}
	 &&\phi'' + \phi'\frac{r'}{r}  = V_\phi,	\\
&&                 \label{EE00}
 	  -\frac12 f(R) +3 H^2 f_R + f_R'' +\frac{r'}{r} f_R' = m_D^{-2}\left( -\frac{\phi'^2}{2} - V \right),
\\&&
                 \label{EE44}
          -\frac12 f(R) - \frac{r''}{r} f_R +\frac{r'}{r} f_R' = m_D^{-2} \left( \frac{\phi'^2}{2} - V \right),
\\&&
		      \label{EE55}
	 -\frac12 f(R) - \frac{r''}{r} f_R + f_R'' = m_D^{-2}\left( -\frac{{\phi'}^2}{2} - V \right),
\ear
  where $f'_R = df_R/du$, etc.

\section{Models with inhomogeneous extra space}

\subsection{Equations and boundary conditions}\label{eqs}

  In our calculations, in order to avoid dealing with third- and fourth-order derivatives, it will be convenient,
  within the same set of equations, to use the Ricci scalar $R(u)$ as one more unknown function. in addition
  to $r(u)$ and $\phi(u)$. As three independent equations for this system,
  we can take, for example, \eqref{phi''}, \eqref{R2} and a combination of \eqref{EE55} and \eqref{R2}:
\bear       \label{phi_2}
	 &&\phi'' + \phi'\frac{r'}{r}  =  V_\phi,	
\\&&            \label{beta_2}
	 R = 12 H^2 - 2\frac{r''}{r},
\\&&             \label{EE5}	
	 -\frac12 f(R) +f_R'' +\left( \frac{R}{2} -6H^2\right) f_R = m_D^{-2}\left(- \frac{\phi'^2}{2} - V \right),
\ear
  resolved with respect to the higher derivatives $\phi'', r'', R''$.
  We will also use the combination \eqref{EE44} + \eqref{EE55} -- \eqref{EE00} -- $f_R\cdot \eqref{R2}$, which leads to
\bear \label{cons2}
	-\frac{f(R)}{2} +  ( R -15H^2) f_R =  m_D^{-2}\left( \frac{\phi'^2}{2} - V \right),
\ear
  and contains lower-order derivatives, as a restriction on the solutions of the coupled
  second order differential equations.

  As boundary conditions, we use the requirement of $u =0$ being a regular center on the
  ($u, \phi$) surface and the corresponding requirements for $\phi$ and $R$:
\bear          \label{ini2}
	&&r(0) =0, \qquad  r'(0) = 1
\\&&		    \label{ini3}
 	\phi(0) = \phi_0, \ \ \ \phi'(0) =0, \ \ \ R(0) =R_0,
\ear
  where all quantities with the index $0$ are constants.
 These initial parameters are related by the condition following from \eq \eqref{cons2},
 \bear                      \label{cons20}
	-\frac{f(R_0)}{2} +  \left( R_0 -15H^2\right) f_R(R_0) = - m_D^{-2} V(\phi_0).
\ear
  This means that for given $f(R)$ the quantity $R_0$ is related to $H$ and $\phi_0$, so that
  any two of these three parameters are free.

  We also have from \eqref{EE5} and \eqref{ini2} for $u \to 0$
\beq             \label{incon_1_}
		R'(0) =0, \quad \ \lim_{u \to  0} \frac{\phi'}{r} = \phi''_0, \quad \ r''(0) = 0.
\eeq

   We will seek solutions for $u > 0$ in which the circular radius $r \to 0$ at some $u = u_{\max}$,
   which provides compactness of the extra space parametrized by $u$ and $\varphi$.

  The total energy on the $(u, \varphi$) surface for a specific solution is
\beq                  \label{rho}
              \rho (\phi_0)= 2\pi\int_0^{u_{\max}} du \ r(u)\left[\frac{\phi'^2}{2} + V\right],
\eeq
  it may be interpreted as the energy density of the scalar field stored in the extra dimensions.
  This energy density depends on the parameter $\phi_0$ expressing the boundary scalar field value in $\M_2$.


\subsection{Pure gravity }

  Let us first consider the case $\phi = \phi_0$, in which the scalar field is distributed uniformly
  in space and does not depend on time, and the equations can be solved analytically.
  In this case the scalar field potential is constant, $V = V_0 = V(\phi_0)$.

 Equation \eqref{cons2} in this case leads to
\beq\label{scaleq0}
		\frac{1}{2} f(R) +(15 H^2-R) f_R = 0,
\eeq
  which means that also $R = R_0 = \const$, hence the 2D extra space is maximally symmetric
  for any given $f(R)$, and from \rf{R2} it follows
\beq
		\frac {r''}{r} = 6H^2 - \frac {R_0}{2}.
\eeq
   Now, the difference of equations (\ref{EE00})--(\ref{EE55}) reduces to
\beq                                     \label{EE05}
		3H^2 f_R + \frac {r'}{r} f_R' + \frac{r''}{r} f_R =0.
\eeq		
  If we assume that $f_R(R_0) \neq 0$, and also notice that $f_R' =0$ due to $R = \const$,
  \eq \rf{EE05} reduces to
\beq                  \label{sin}
			 \frac{r''}{r} = -3 H^2,
\eeq
  and we also have
\beq \label{R4}
		 R = R_0 = 18 H^2,    \qquad      \frac{f}{f_R}=6 H^2.
\eeq

   Under our conditions at $u=0$, the solution of \rf{sin} reads
\beq
		 r  = \frac{1}{\sqrt{3} H} \sin\Big(\sqrt{3} H \, u\Big),
\eeq
  and the metric has the form
\bear                    \label{ds4}
		ds^2 &=& dt^2 -\e^{2 H t} \delta_{i j} \ dx^i dx^j  - \frac{1}{3 H^2}
		\Big( d \theta^2 +  \sin^2 \theta d\varphi^2 \Big), \nn
        &&\theta = \sqrt{3} H\, u,
\ear
  the extra space being a 2-sphere.

  We have shown that with any choice of the initial function $f(R)$ the only solution with the
  metric \rf{ds2} for pure gravity (or with a constant scalar field) corresponds to a spherical
  extra space.

  This result deserves attention at high energies where the Hubble parameter is large enough. A common
  starting point is to fix the properties of extra dimensions, their size in particular. These
  properties depend on the Lagrangian parameters, including the topology of extra space, but do not depend
  on our 4D metric. According to \eqref{R4}, \eqref{ds4}, the state of affairs is different at least
  for the class of models containing all sorts of $f(R)$. The extra space is inevitably maximally
  symmetric, and its radius is stiffly related to the Hubble constant, $r=\sqrt{3}H^{-1}$.

  In particular, if we choose $f(R) = a R^2 +R +(c-V)$ (see \eq \rf{f(R)} further on),
  \eq \eqref{scaleq0} gives the following relation between the parameters:
\bear
            &&108 a H^4 +12 H^2 +c-V=0 \quad \ \mbox{or} \nn
            &&H^2 = \frac{-b \pm\sqrt{b^2 -3 a (c-V)}}{18 a} = \frac{R_0}{18}.
\ear
  The possibility of complex roots in this expression shows that not any choice of the parameters leads
  to a valid solution, since obviously $H^2$ must be real.

  At the inflationary stage of the Universe evolution, $r$ is close to $10^{-27}$ cm, and it is about
  $10^{-33}$ cm at the Planck scale. However, if we consider very small $H$, for example,
  corresponding  to the present epoch, the Ricci scalar of extra dimensions will be close to
  zero, meaning their huge size, incompatible with observations. To avoid such a strong constraint,
  one can add matter fields (a scalar one in our case) or/and widen the Lagrangian by adding other
  invariants like the Ricci tensor squared, making it possible
  to obtain inhomogeneous extra dimensions. A detailed discussion on the basis of other extra space
  metrics can be found in \cite{2018EPJC...78..373P, 2010IJGMM..07..797I}.

\subsection {Numerical solutions. Conical singularities.}

  To obtain examples of numerical solutions of interest, let us choose the following functions in
  the action \rf{S}:
\bear                           \label{f(R)}
		&&f(R)=aR^2 +bR +c, \qquad a,b,c = \const,
\nn &&                               \label{V}
		V(\phi)=\fract{m^2}{2} \, \phi^2,	\qquad  m = \const.	
\ear

  The figures below present solutions for different values of the parameters.
\begin{figure}[ht!]
\centering
\includegraphics[width=7cm]{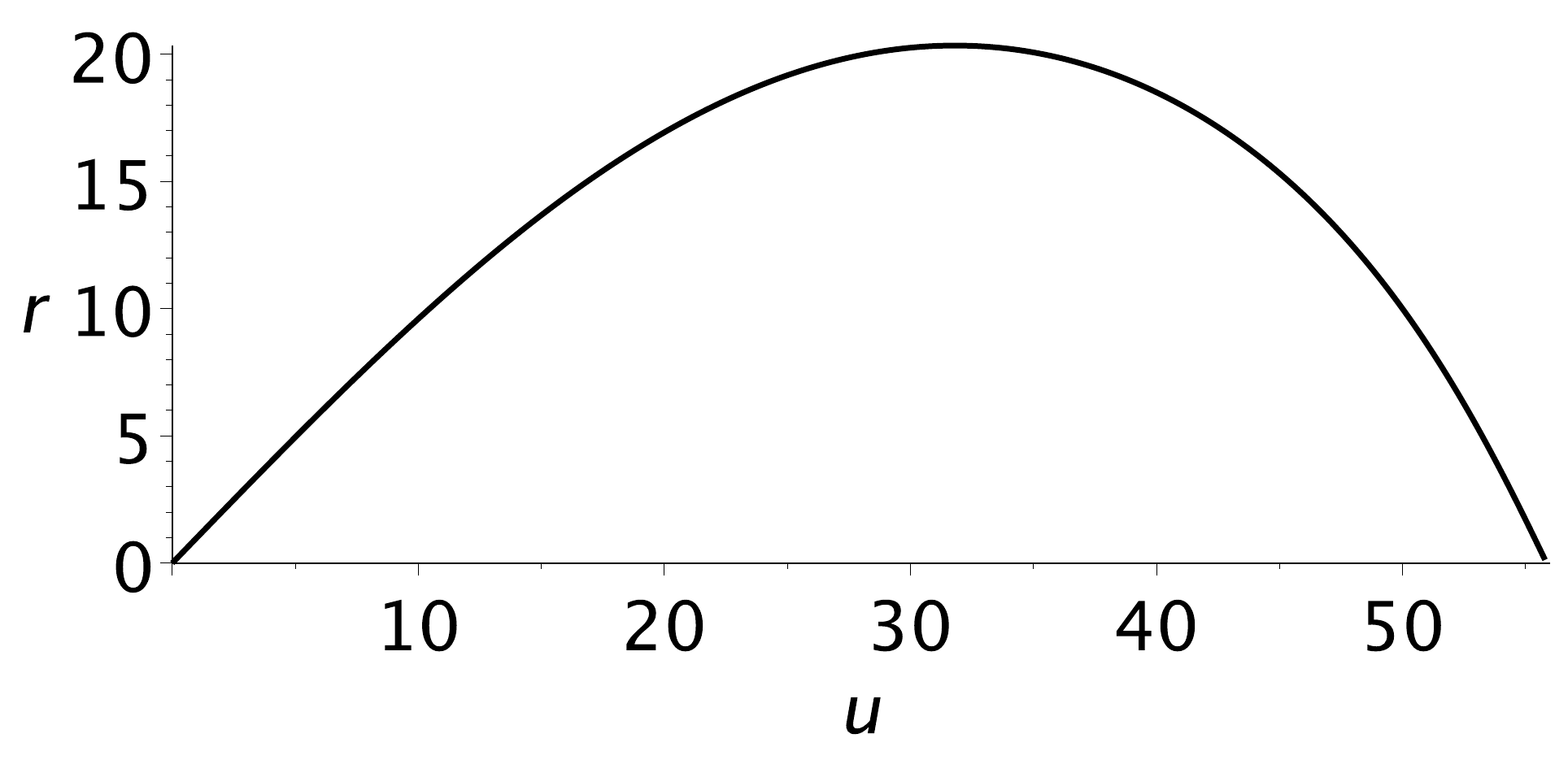} \quad \includegraphics[width=7cm]{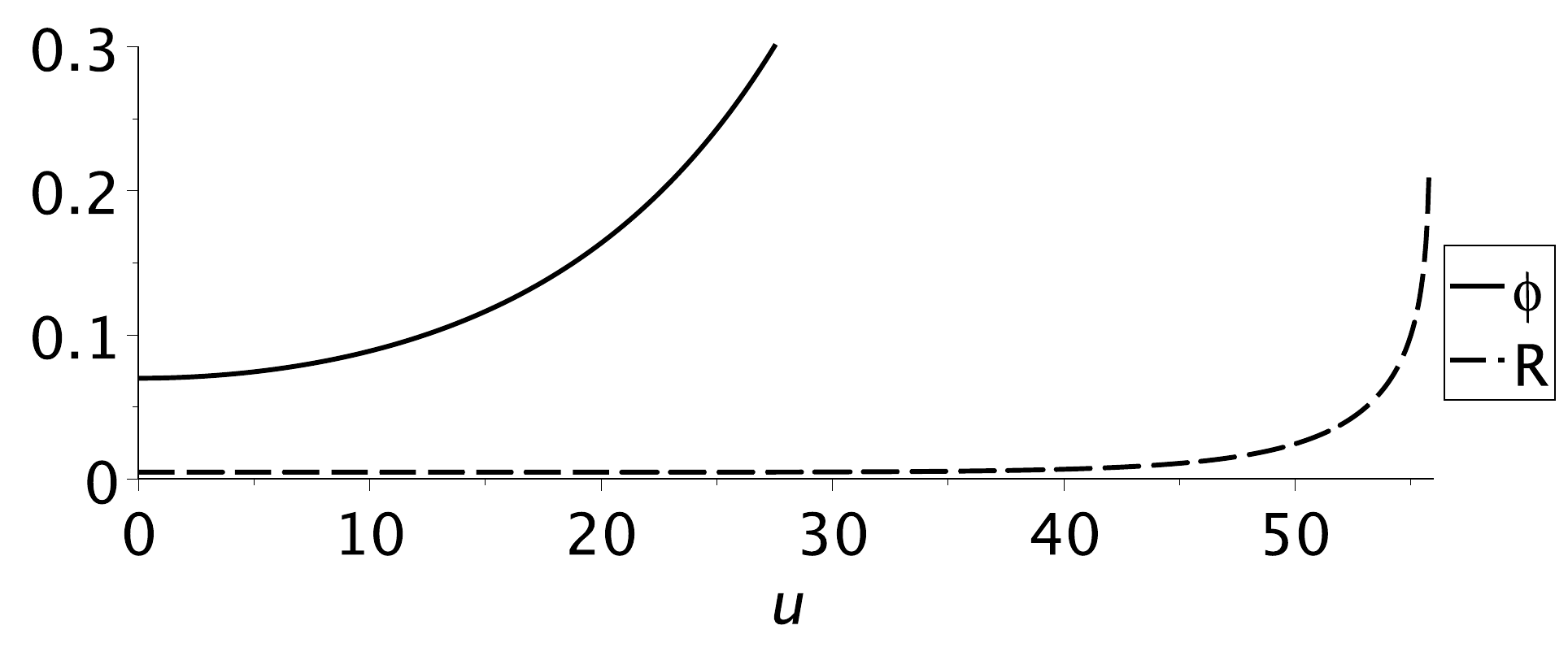}
\caption{\small
	The extra space metric function $r(u)$, the Ricci scalar $R(u)$ of 6D space and the scalar field $\phi(u)$
	for $f(R)=aR^2 +bR +c$ and $V(\phi)=(m^2/2) \phi^2$ (units $m_D=1$). The parameter values are
	$ m=0.1, \ b=1, \ a=-100, \ c=-0.0021$. Additional conditions are: $\phi_0=0.07, \ H=0.$
	The value of $R(0)$ follows from \eq \eqref{cons20}. Here  $R(0) \simeq 0.00485$. }
\label{Heq0} \end{figure}
\begin{figure}[ht!]
\centering
\includegraphics[width=7cm]{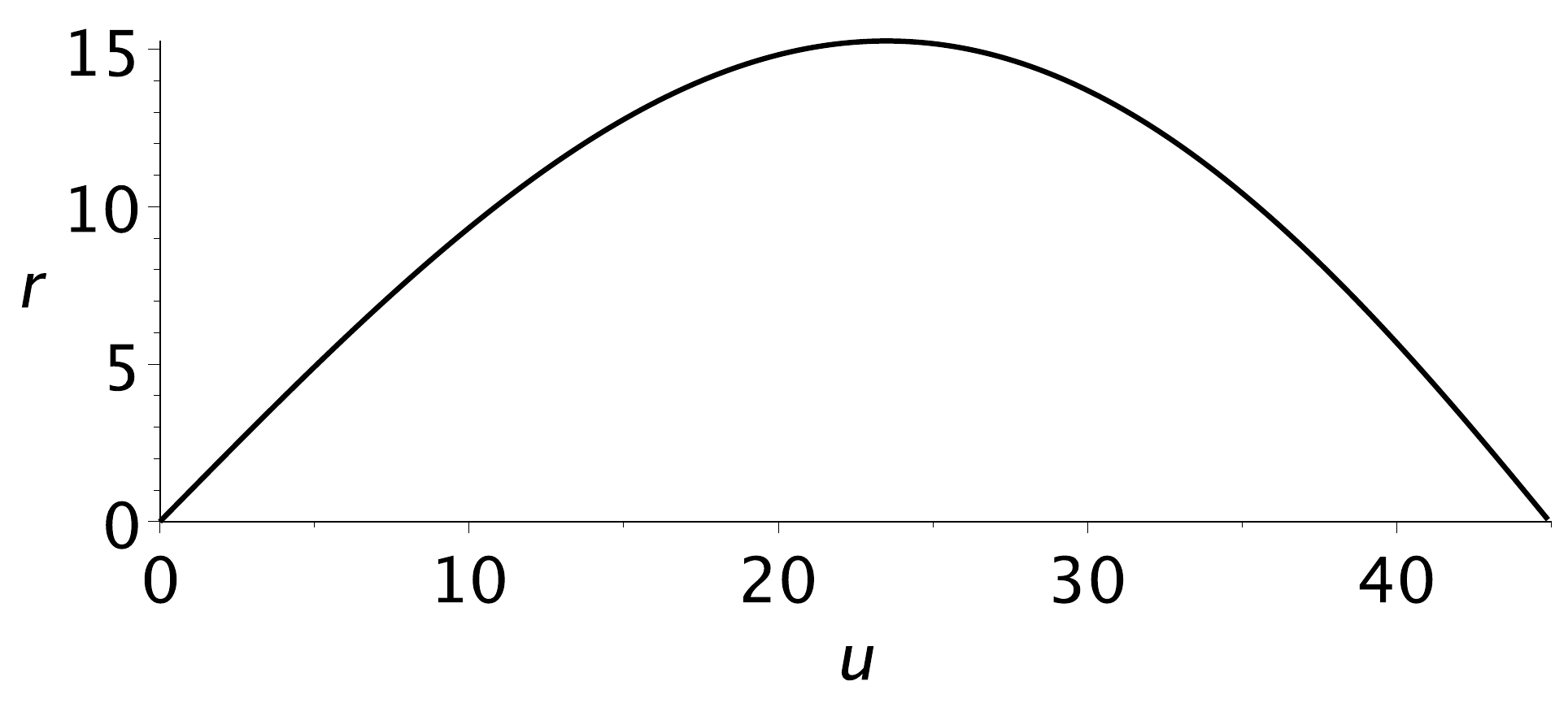} \quad \includegraphics[width=7cm]{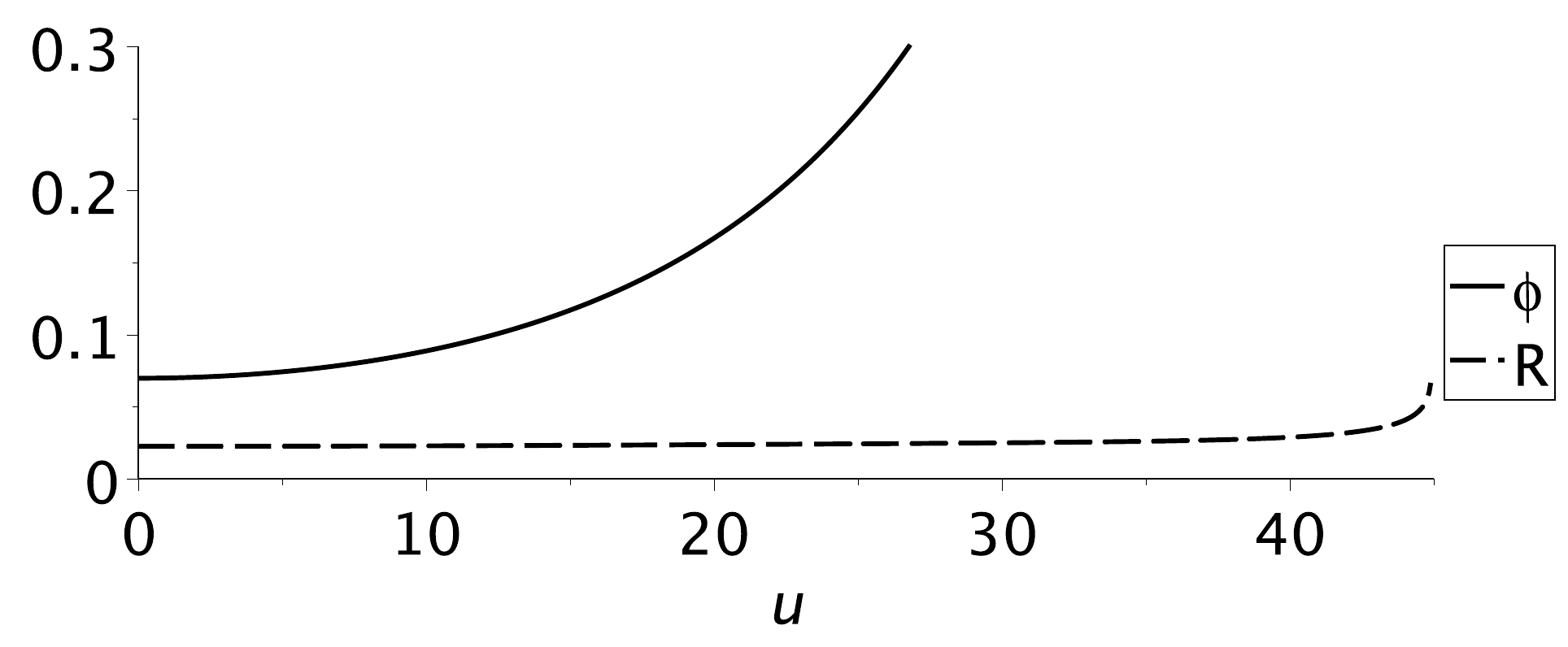}
\caption{\small
    The same as in Fig.\,1 but $H=0.035, \ R(0) \simeq 0.0227767$.}
\label{Heq0035} \end{figure}
\begin{figure}[ht!]
\centering
\includegraphics[width=7cm]{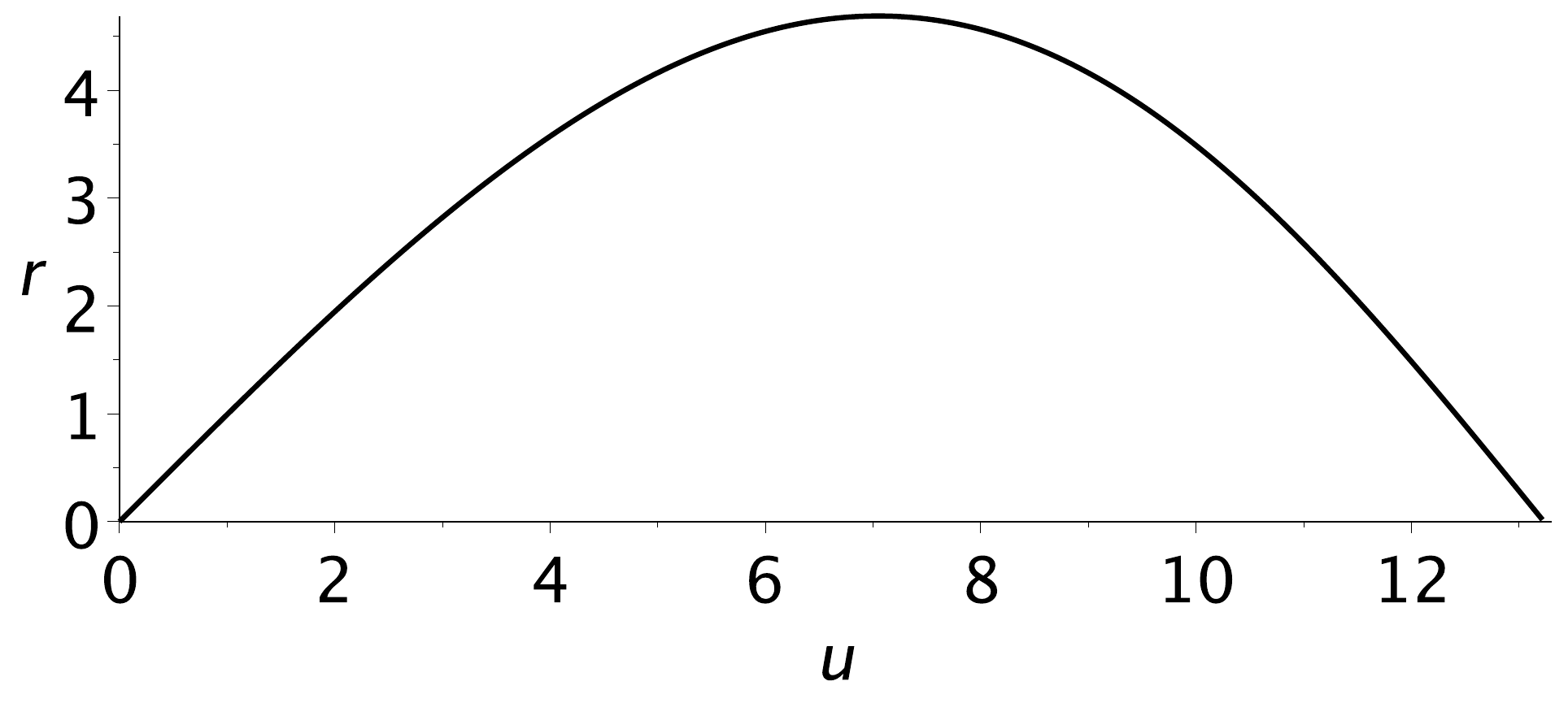} \quad \includegraphics[width=7cm]{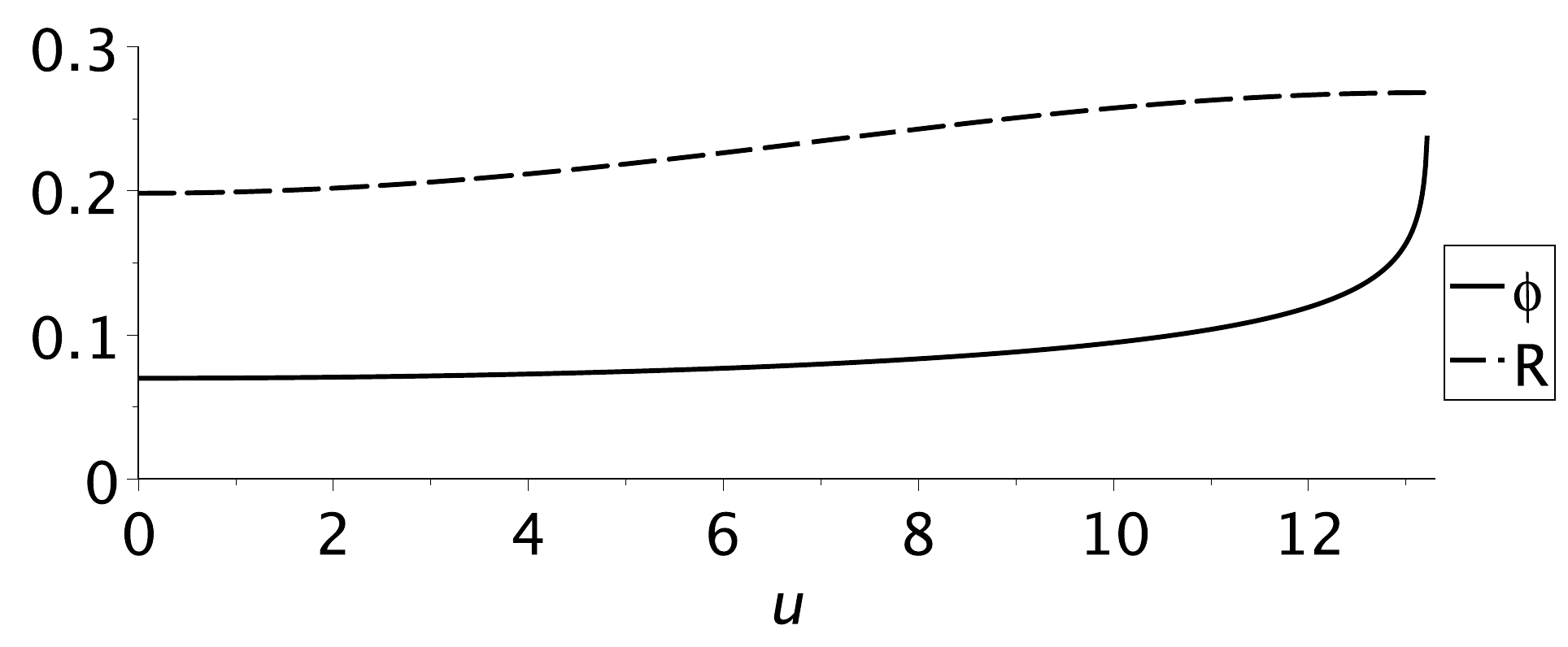}
\caption{\small
	The same as in Fig.\,1 but $H=0.1, R(0) \simeq 0.198327 $.
}
\label{Heq01} \end{figure}
\begin{figure}[ht!]
\centering
\includegraphics[width=7cm]{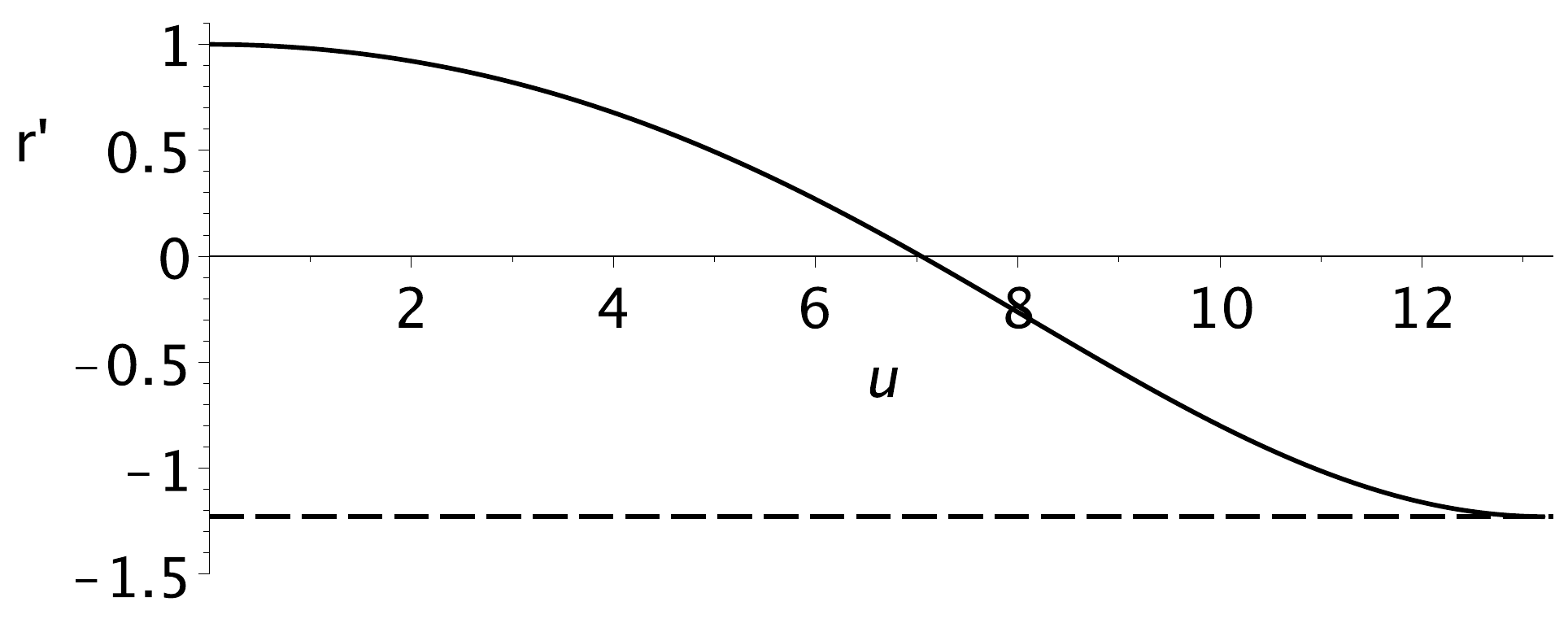} \quad \includegraphics[width=7cm]{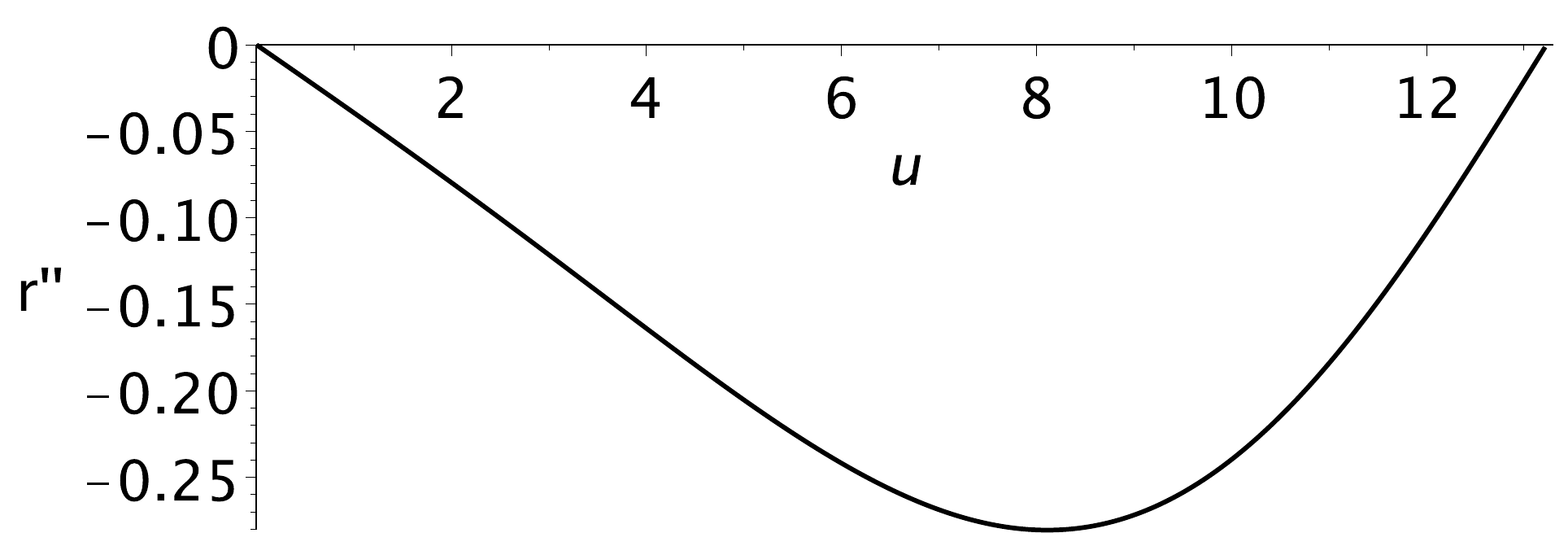}
\caption{\small
	Detailed information about the conical singularity shown in Fig.\,3. The first derivative $r'(u_{max})<-1$ while the second derivative $r(u_{max})=0$.
}
\label{Heq01a}\end{figure}
\begin{figure}[ht!]
\centering
\includegraphics[width=7cm]{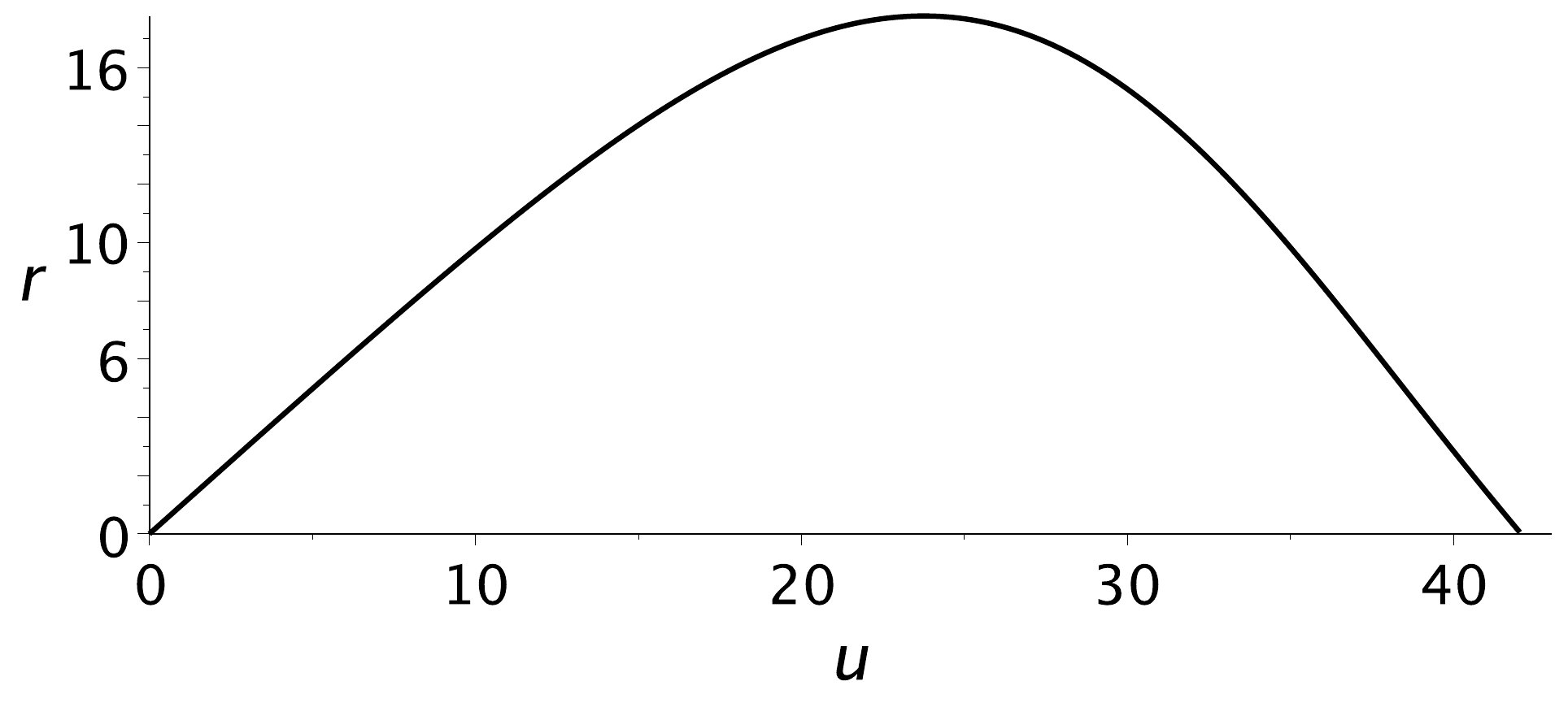} \quad \includegraphics[width=7cm]{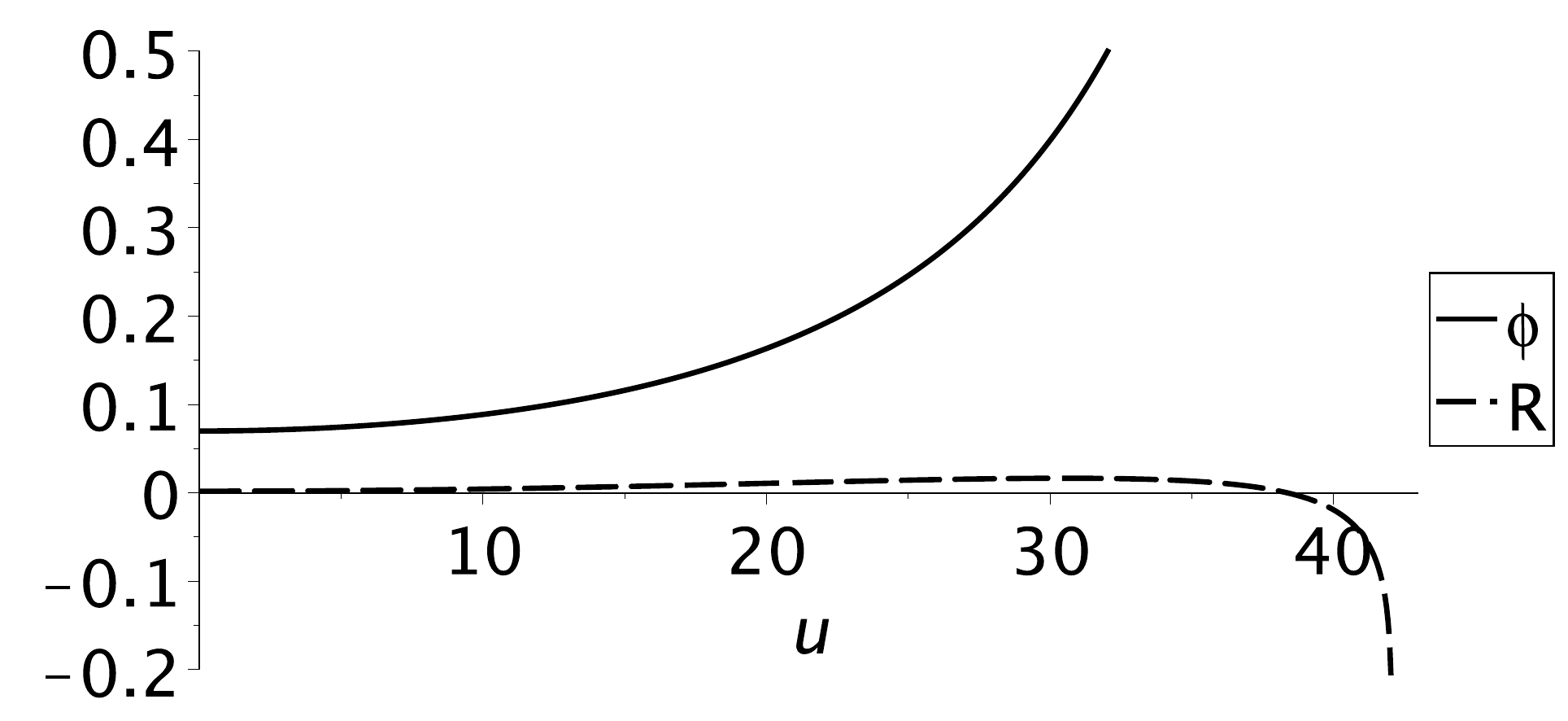}
\caption{\small
      The parameter values: $ m=0.1, \ b=1, \ a=10, c=0.0021$. Additional conditions:
      $ \phi_0=0.07, \ H=0, \ R(0) \simeq 0.00189$ follows from \eq \eqref{cons20}.}
\label{fig7}\end{figure}

  The numerical results are presented in Figs.\,1--6.
%
%
  A variation of the parameter values can lead to qualitatively different metrics in $\M_2$.
  For example, the curvature may change its sign, as is seen from Fig.\,\ref{fig7}.

\begin{figure}[ht!]
\centering
\includegraphics[width=5cm]{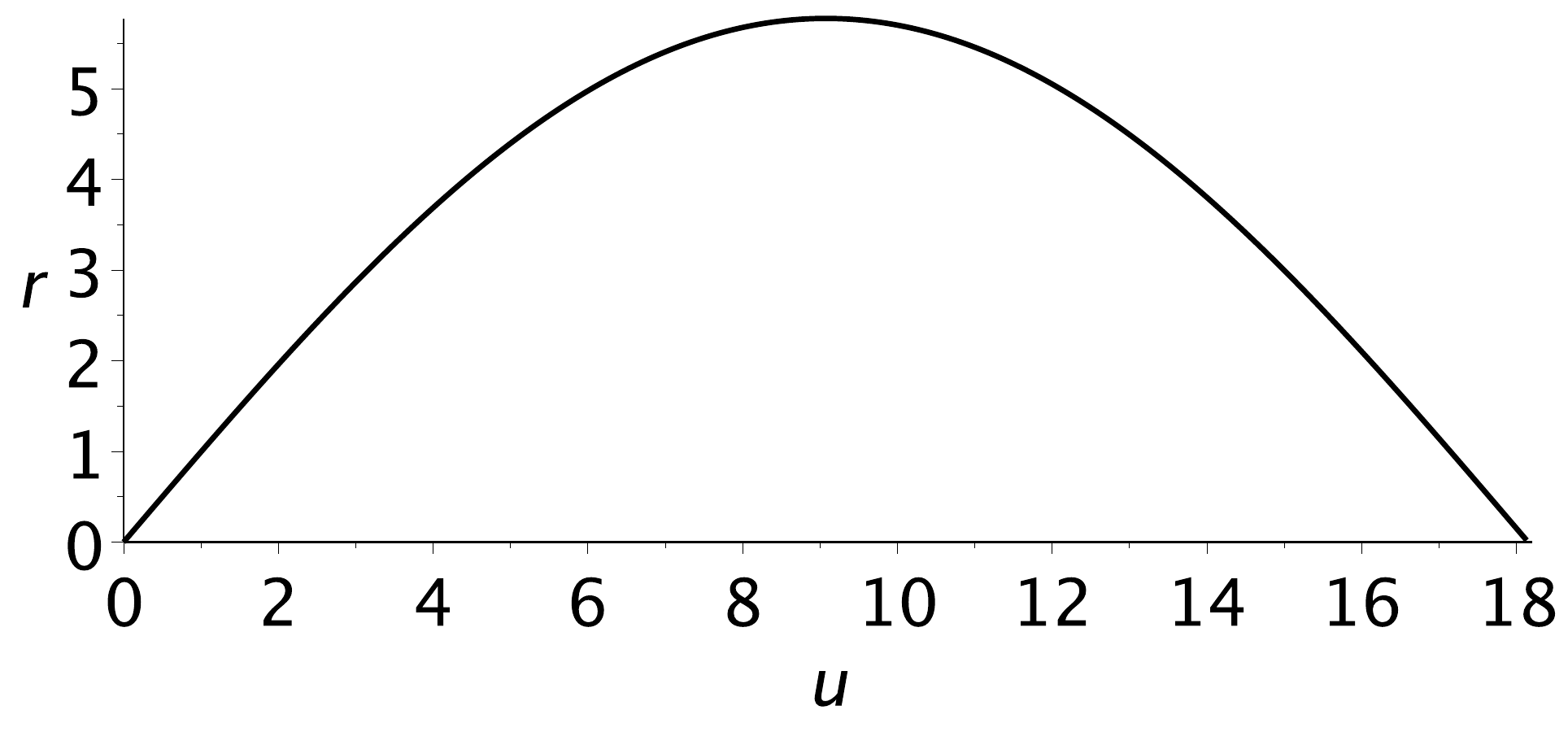} \  \includegraphics[width=5cm]{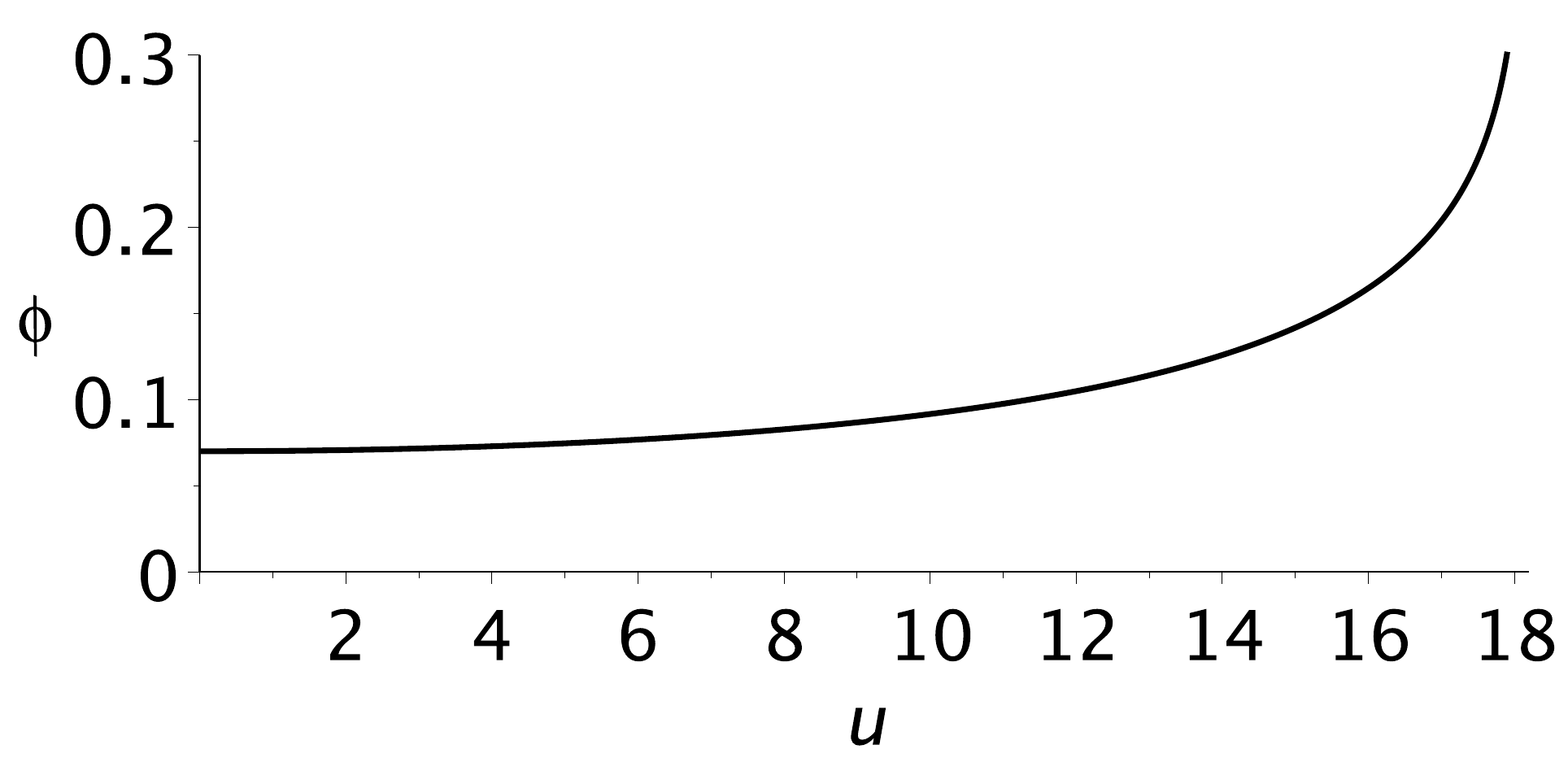} \  \includegraphics[width=5cm]{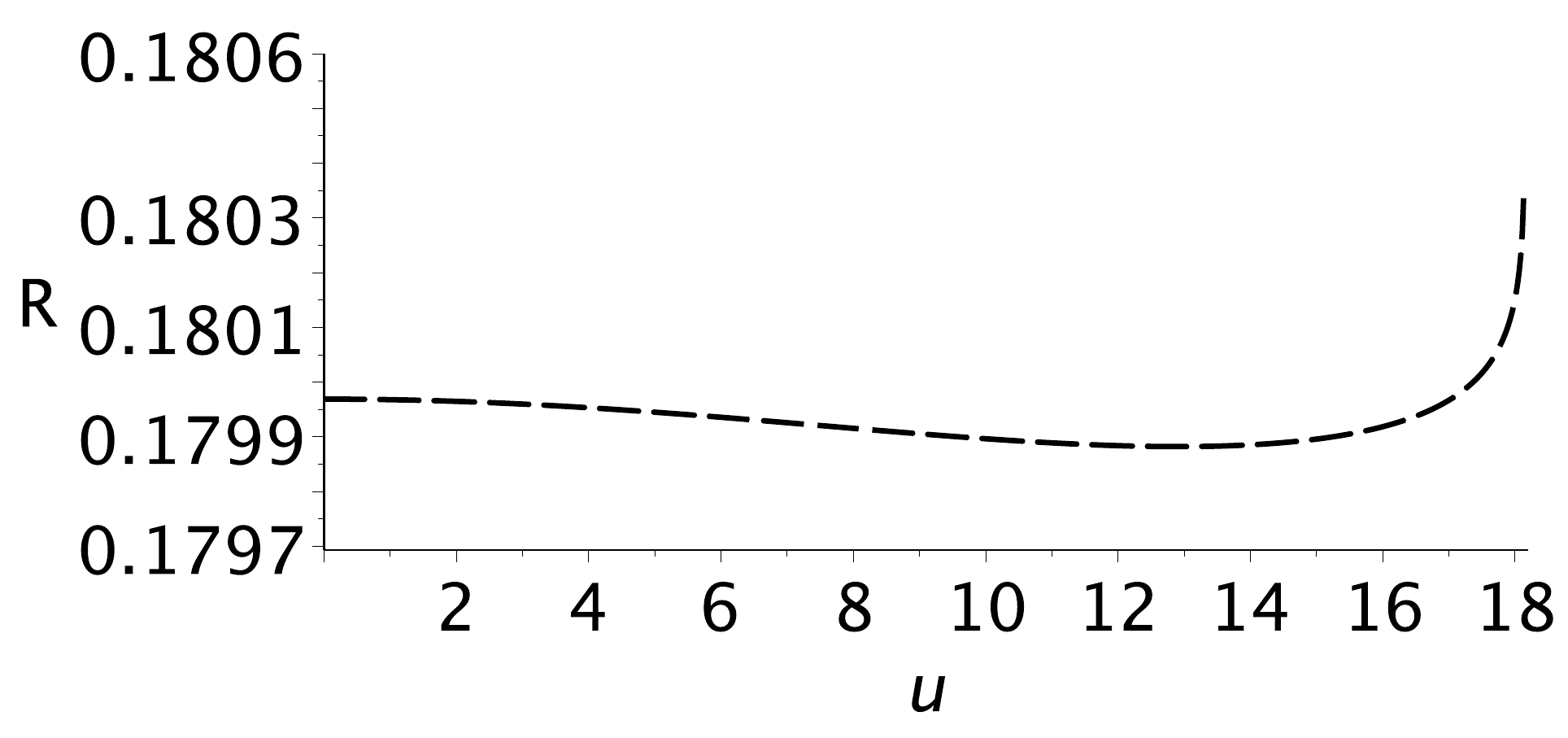}
\caption{\small Solution without a conical singularity, $ r'(u_{\max}) = -1$.
	The extra space metric function $r(u)$, the Ricci scalar $R(u)$ of 6D space and the scalar field $\phi(u)$
	for $f(R)=aR^2 +bR +c$ and $V(\phi)=(m^2/2) \phi^2$ (units $m_D=1$). The parameter values are
	$ m=0.1, \ b=1, \ a=-10.9, \ c=-0.0021$. Additional conditions are: $\phi_0=0.07, \ H=0.1$
	The value of $R(0)$ follows from \eq \eqref{cons20}. Here  $R(0) \simeq 0.179969, r'(u_{\max}) = -1 $.}
\label{sol}
\end{figure}

  The point $u=0$ is a regular center by our boundary conditions. One can notice that in general our solutions contain conical
  singularities at $u=u_{\max}$.
  Indeed, as is seen from Fig. \ref{Heq01a}, left panel, where the first derivative $r'(u_{\max})\neq -1$, the curvature $R$
  is infinite at the point $u=u_{\max}$. It is a pointwise singularity completely similar to that of a usual top of a cone:
  the limiting values of $R$ as $u \to u_{\max}$ are finite, see Fig.\,\ref{Heq01}, right panel. Moreover, the l.h.s. in
  \eqn{cons2} is finite in the same limit, so finite is also the r.h.s. characterizing the
  scalar field.  Such metrics are used in the extra dimensional context, see, e.g., \cite{Gogberashvili_2007}
  and references therein.

  It is also of interest to check whether or not there are solutions free from such a singularity, other than the evident
  maximally symmetric metric, $R= \const$. The question is: can we find a nontrivial nonsingular solution? The answer is yes,
  and it is presented in Fig.\,\ref{sol}. To be sure that this is not a numerical effect, we shifted the value
  $\phi_0=0.07$ in both directions and accordingly obtained $r'(u_{\max}) < -1$ and $r'(u_{\max}) > -1$.

  In this solution with a regular metric, as can be verified through the field equations, the
  curvature $R$ and the scalar field are also finite, so such models are completely regular. However, models
  with conical singularities do not differ too substantially from these regular models in their physical properties.

  A few general remarks on possible singularities in extra dimensions.
  Evidently, the classical equations written above are invalid at energy scales larger than the
  Planck scale. For 4D physics the corresponding length scale is about $l_4=1/m_4\sim 10^{-33}$ cm.
  At scales near $l_4$ and smaller, quantum fluctuations are strong, and any solution to the classical
  equations is invalid. In a $D$-dimensional world a similar scale is $l_D=1/m_D=1$ by our convention.
  There are two consequences if we intend to work on the classical level: (i) the size of extra
  dimensions must be much larger than unity; (ii) any peculiarities with the coordinate interval
  $\delta u \simeq l_D=1$ are meaningless without thorough analysis of quantum effects. In particular, if
  a classical solution contains a singularity, as it happens in most of our solutions at $u = u_{\max}$, it
  is reasonable to suppose that such a singularity is smoothed by quantum effects and should not be
  taken seriously. The stability of such configurations based on the methods discussed in \cite{Zhong_2016}
  is considered below.

\section{Stability}

  A general analysis of the stability is a very complicated task. Here we show that our solutions are
  stable relative to perturbations homogeneous in 4D space, depending only on $t$ and $u$.
  More definitely, we study the evolution of metric \eqref{ds2} with the small deviations
\bear                \label{delta2}
	&&\phi(t,u)=\phi_c (u)+  \delta\phi(t,u),\
\nn &&	
	r(t,u)=r_c(u)+\delta r(t,u), \quad   R(t,u)=R_c(u) +\delta R(t,u),
\nn &&
	\delta\phi \ll \phi_c,\quad \delta r \ll r_c, \quad \delta H \ll H_c, \quad \delta R \ll R_c.
\ear
Here and below the index $"c"$ relates to the static solutions.
  We substitute them into the classical equations and show that there are no growing modes.
  There are three unknown functions $\delta r(t,u)$, $\delta \phi(t,u)$, $\delta R(t,u)$, so that we need
  three classical equations linearized with respect to these quantities, which may be written in the form
\bear             \label{dphi}
	&&-\Box_2 \delta\phi -3H_c \delta {\dot \phi}
	+\frac{1}{r_c} r'_c \delta \phi'  + \frac{1}{r_c }\phi'_c \delta r'
	-\frac{1}{r_c^2 }\phi'_c r'_c \,\delta r
\nn &&
     \hskip55mm -m^2 \delta\phi =0,
\\&& 			 \label{dR0}
	\delta R = \frac{2}{r_c} \Box_2 \delta r  +6 \frac{H_c}{r_c} \delta {\dot r} +\frac{2}{r_c^2} r''_c \,\delta r,
\ear
\bear\label{dtrace}
		&&a \bigg[\frac{4R_c}{r_c} \Box_2 \delta r - 10 \Box_2 \delta R
		+12 \frac{H_c R_c}{r_c} \delta {\dot r}  +\frac{10 R_c'}{r_c} \delta r'
\nn &&
		- 30 H_c  \delta {\dot R} + \frac{10 r_c'}{r_c} \delta r'
        +\frac{1}{r_c^2}\Big(4r_c'' R_c -10 r_c' R_c'\Big) \delta r
\nn &&
		+\frac{1}{r_c^2}\Big(4r_c'' R_c -10 r_c' R_c'\Big) \delta r
		+\left( 24 H_c^2 -\frac{4 r_c''}{r_c} - 6R_c \right) \delta R \bigg]
\nn &&	
		+b \bigg[ \frac{2}{r_c} \Box_2 \delta r
		+\frac{6 H_c}{r_c} \delta {\dot r} +\frac{2 r_c''}{r_c^2} \delta r - 3 \delta R\bigg]
		+ 6 m^2 \phi_c \delta \phi
\nn &&
        +4 \phi_c' \delta \phi'=0,
\ear
  where the dot denotes $\d/\d t$, $\Box_2 = \d_{tt}^2-\d_{uu}^2$, $a$ and $b$ are coefficients from
  \eqref{f(R)}.

  In fact, we arbitrarily choose initial deviations from the static field configuration and perform
  numerical analysis. We show that the solutions to the classical equations have no growing modes and relax
  (due to friction) to the static homogeneous configuration. The friction is supplied by the nonzero Hubble
  parameter $H$ and (as confirmed by calculations) is absent if $H=0$.

  For the perturbation equations we have used the boundary conditions
\bear
             &&\delta r(t,0)=\delta \phi (t,0)=\delta R(t,0)= \delta r(t,u_{\rm max})
             \nn &&
             =\delta \phi (t,u_{\rm max})=\delta R(t,u_{\rm max})=0
\ear
  and the initial conditions
\bear
    &&\delta r(0,u)=\delta \phi (0,u)=\delta R(0,u)= 0.01 \sin^2{(u\pi/u_{\rm max})}
\nn &&
     \delta {\dot r}(0,u)  = \delta {\dot \phi} (0,u) = \delta R(0,u)=0.
\ear
\begin{figure}[ht!]
\begin{center}
\begin{overpic}[width=8cm]{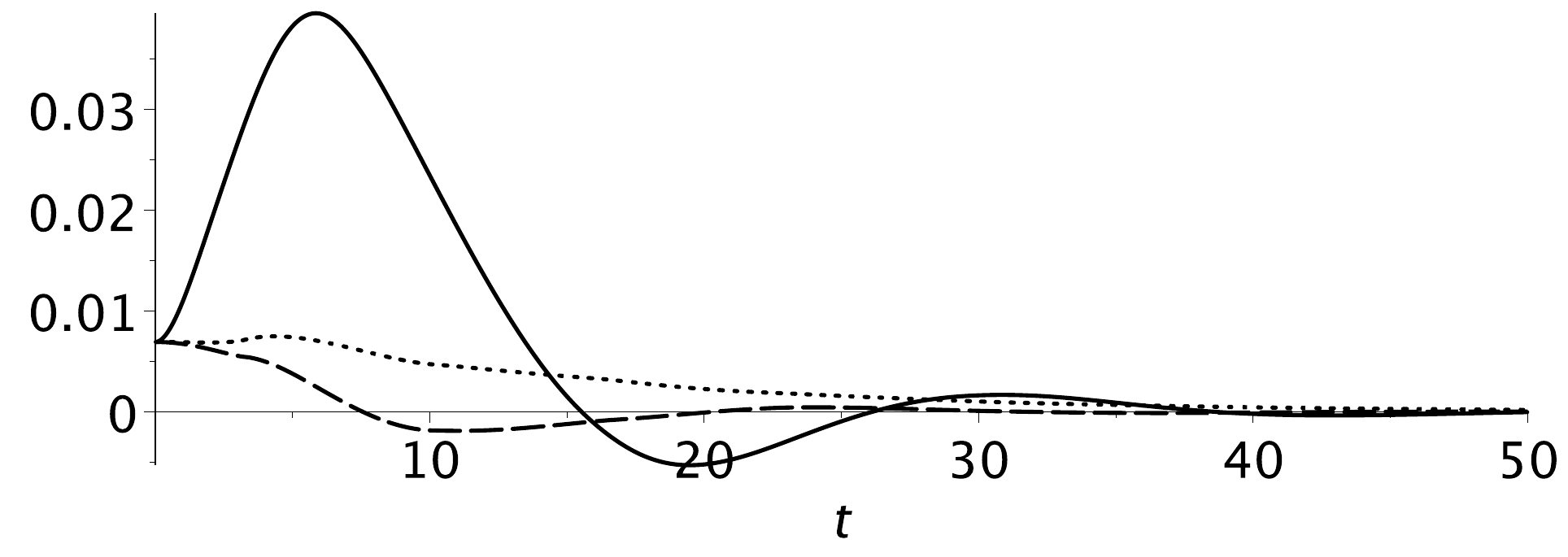}
\put(60,25){   \textrm{ {\bf ---} $ \delta r(t, u=10)$}}
\put(60,20){   \textrm{ $ \cdots \ \delta \phi(t, u=10)$}}
\put(60,15){   \textrm{ ${\bf --}\ \delta R(t, u=10)$}}
\end{overpic}
\caption{\small Time dependence of the perturbations to the solution presented in Fig.\,\ref{Heq01}}
\end{center}
\label{perts}
\end{figure}

\begin{figure}[ht!]
\begin{center}
\begin{overpic}[width=8cm]{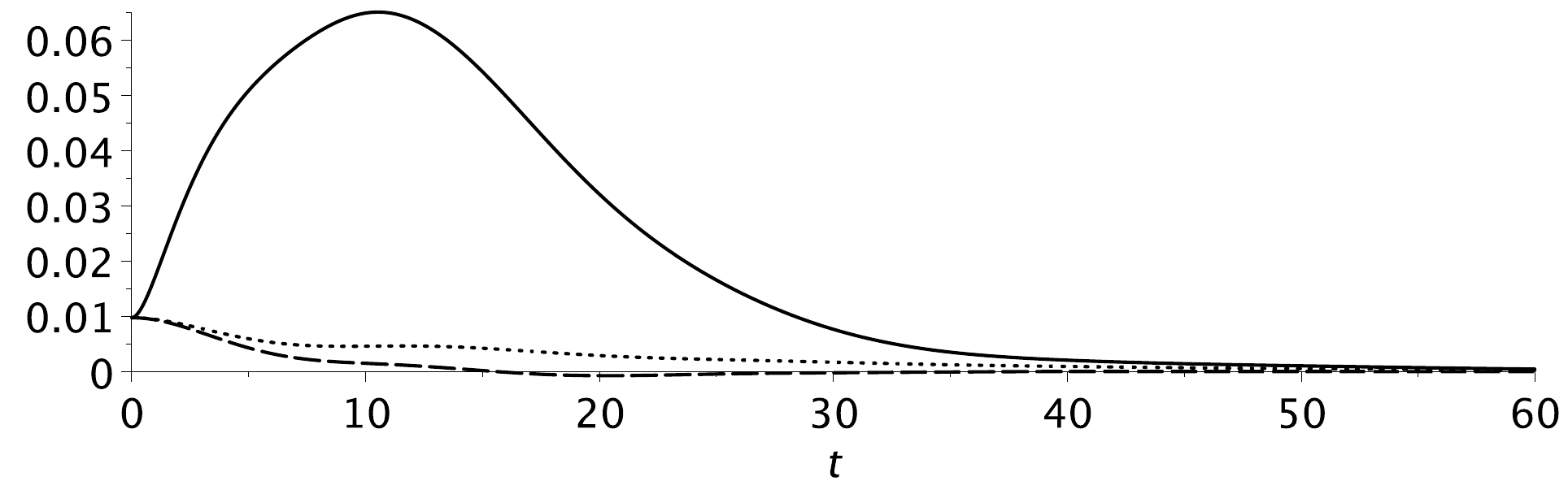}
\put(60,25){   \textrm{ {\bf ---} $ \delta r(t, u=10)$}}
\put(60,20){   \textrm{ $ \cdots \ \delta \phi(t, u=10)$}}
\put(60,15){   \textrm{ ${\bf --}\ \delta R(t, u=10)$}}
\end{overpic}
\caption{\small Time dependence of the perturbations to the solution presented in Fig.\,\ref{sol} (values at $u=10$)}  \label{perts2}
\end{center}
\end{figure}

  The nonsingular solution presented in Fig.\ref{sol} is also stable. The time behavior of its fluctuations under
  the same boundary and initial conditions is shown in Fig.\,\ref{perts2}. 


\section{Reduction to 4 dimensions and low energies}

  The study made above reveals that static inhomogeneous extra dimensions could exist.
  For given $f(R)$ and $V(\phi)$, their shape and energy density also depend on the
  initial (random)l value $\phi_0$. Let us briefly discuss the observational manifestations
  of such solutions.  As was shown in \cite{Lyakhova:2018zsr}, there exist such extra-dimensional metrics that
  lead to the 4D cosmological constant $\Lambda_4$ arbitrarily close to zero. This effect is a result
  of interference between the gravitational and scalar field parts of the Lagrangian. The result
  obtained in \cite{Rubin:2015pqa} is based on approximate equations. In this section, we use the exact set
  of equations derived from the metric \eqref{ds2} and the action \eqref{S}.

  The quantity $\Lambda_4$ can be found by integrating out the internal coordinates in the
  action \eqref{S}. We know that the Hubble parameter is at present almost zero as compared to
  the possible extra-dimensional scales. Hence let us put $H=0$. In this case $R_0$ and $\phi_0$
  are related by \eqref{cons20}, and $\Lambda_4(\phi_0)$ is a function of the unique argument
  $\phi_0$. It remains to find this function and its zero points.

  Let us consider static solutions found above
  and use the smallness of $H$ as compared to the extra space Ricci scalar $R_2=- 2 r''/r$,
  see \eqref{R2}. After the decomposition $f(R)=f(R_4+R_2)\simeq f(R_2)+ f_R(R_2)R_4$ we obtain
\bear
	S&=&2\pi \int d^4 x\sqrt{g_4}\int_0^{u_{\max}}du r(u)
	\left[\frac12 f_R(R_2)R_4 + \frac12 f(R_2)
\right.\nn && \left.
    -\frac12\phi_{,u}^2 -V\big(\phi(u)\big)\right]
\ear
  Comparing this expression with the standard form of the 4D action
\beq                 \label{standd}
    	S_4 = \int d^4 x\sqrt{g_4}\left[\frac12 m_4^2 R_4 -\Lambda(\phi_0)\right]
\eeq
  we get the observed Planck mass
\beq                  \label{m4}
    m_4^2 (\phi_0)=2\pi\int_0^{u_{\max}}du\, r(u)f_R(R_2)
\eeq
  in the units $m_D=1$,  and
\bear \label{Lambda}
             \Lambda_4 (\phi_0)&=&-2\pi\int_0^{u_{\max}}du \, r(u)
             \left[ \frac12 f(R_2) -\frac12\phi_{,u}^2
             \right. \nn && \left.
             -V\big(\phi(u)\big)\right].
\ear
\begin{figure} \label{Lambda2}
\includegraphics[width=7cm]{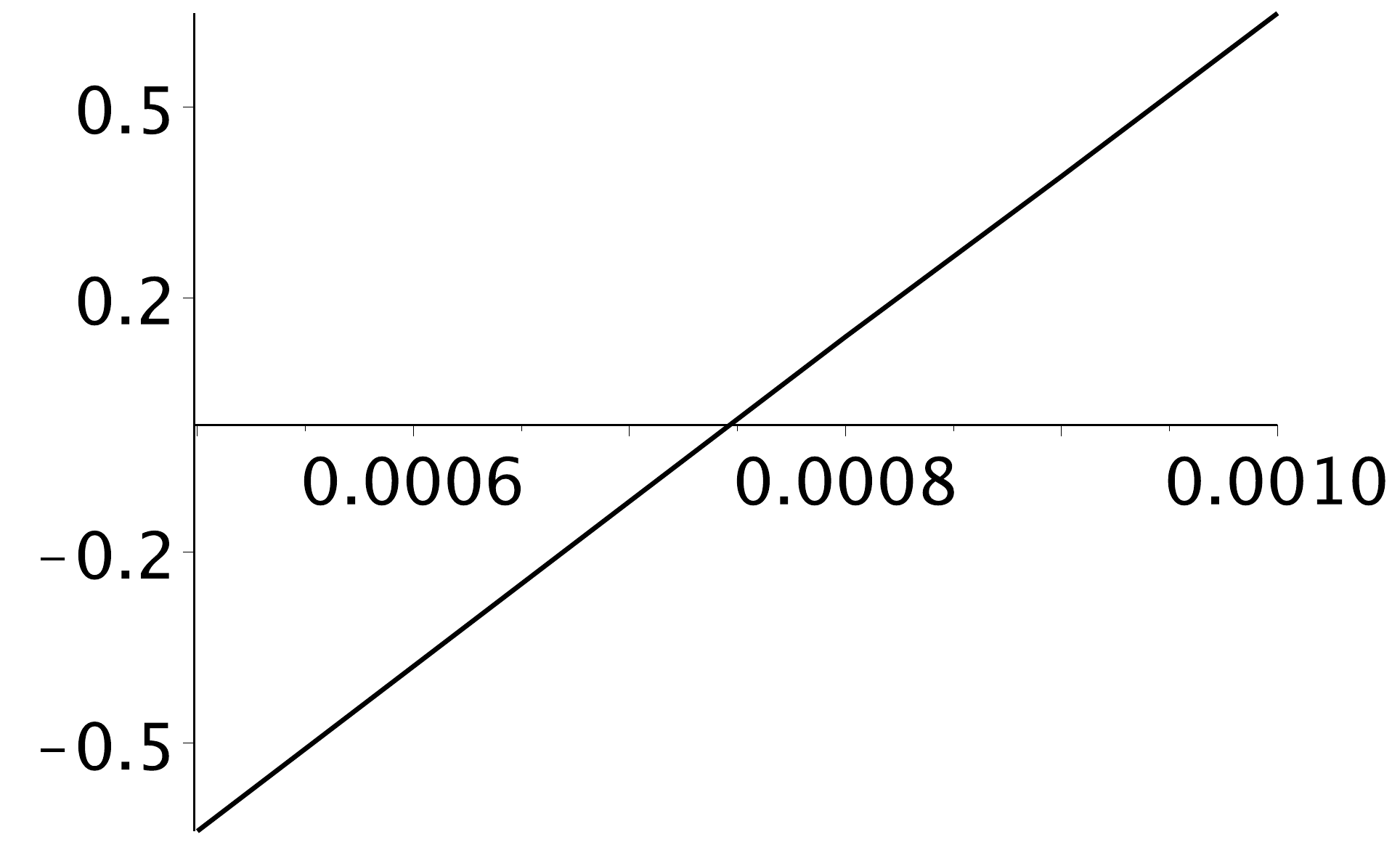} 
\ \includegraphics[width=7cm]{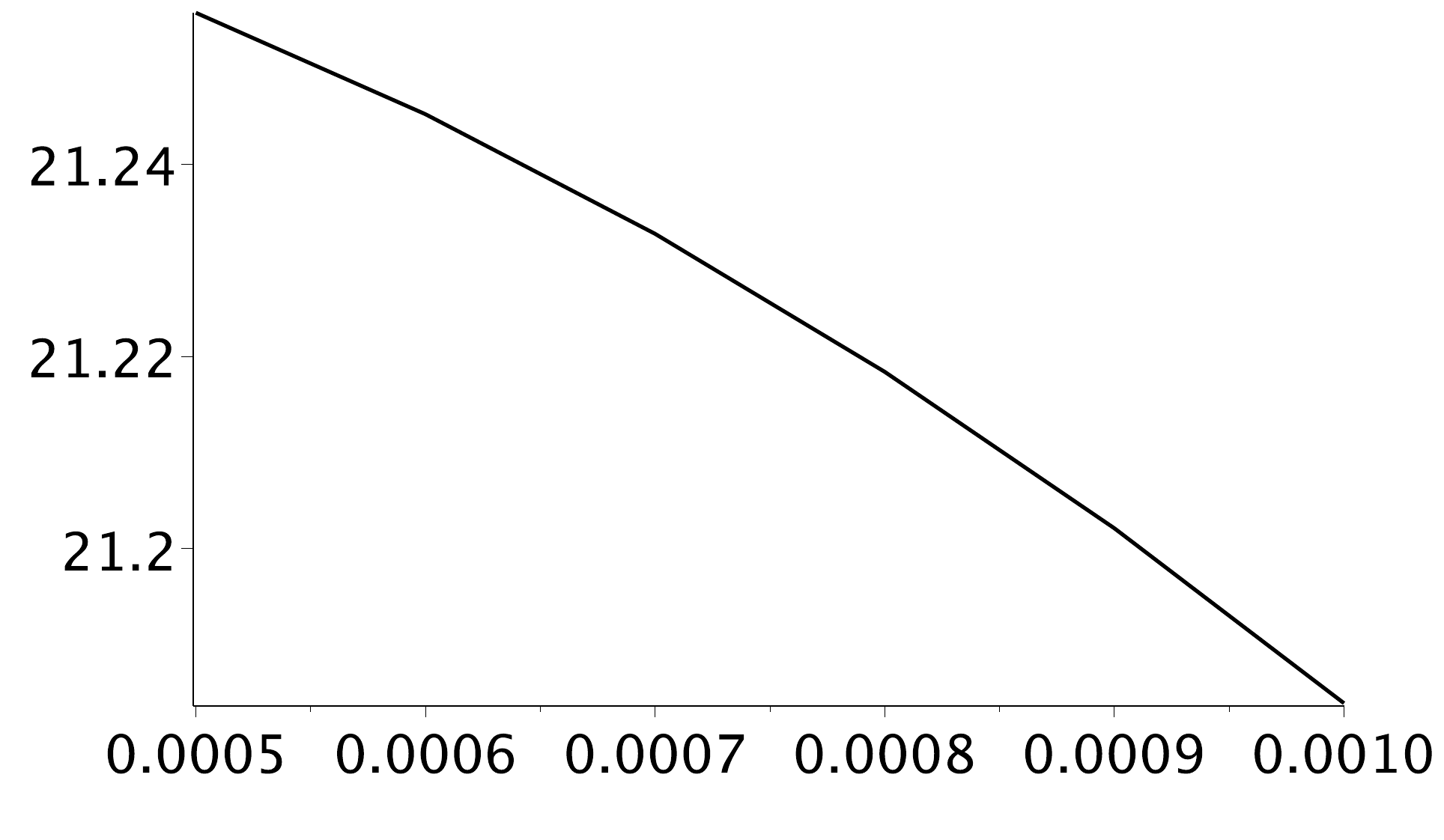}
\put(-400,100){$\Lambda$}
\put(-265,73){$\phi_0$}
\put(-185,110){$m_4$}
\put(-25,33){$\phi_0$}
\caption{\small
	The dependence of $\Lambda_4$ and $m_4$ on $\phi_0$ for $f(R)=aR^2 +bR +c$
	and $m_D=1, \ m=0.1, \ b=1, \ a=-100, \ c=-0.0021, \ V(\phi)=(m^2/2) \, \phi^2, \ H=0$. $R(0)$
	is the positive root of \eq \eqref{cons20}.}
\end{figure}

  One can easily find the value $\phi_0 \simeq 0.00073$ that corresponds to the zero value of
  $\Lambda_4$ from Fig.\,7, left panel. The right panel gives the 4D Planck mass
  $m_4\simeq 21.23$ in units $m_D=1$. Now everything is prepared to calculate the energy
  density \eqref{rho} of the scalar field distributed within the extra dimensions. Numerical integration
  in $u$ gives
\beq                           \label{dens}
	    \rho (\phi_0) \simeq 1.03 \ m_D^4 \simeq 1.03 (m_4 / 21.23)^4  \simeq  0.0000051\, m_4^4.
\eeq
  The scalar field density stored in the extra space is neutralized by the gravitational term $f(R_2)$,
  so that the cosmological constant \eqref{Lambda} is small for the specific solution to \eqs \eqref{eqs}.
  Such a solution certainly exists due to the continuity of the set of solutions. The quantity $\phi_0$
  was used as an additional parameter, see \eqref{ini3}, to find a specific distribution $\phi(u)$.
  We see that the set of the scalar field distributions is parametrized by the boundary value $\phi_0$.
  The same can be said about its energy-momentum tensor $T_{AB}(\phi_0)$.

\section{Conclusion}

  In this paper, we have studied the metric of compact extra dimensions at high energy density of the Universe
  where the 4D space-time is described by the de Sitter metric with an arbitrary value of the Hubble parameter.
  Numerical solutions to the full set of the classical equations have been analyzed. It is shown that in a theory
  with given $f(R)$, inclusion of a scalar field leads to a continuous set of static extra space metrics.
  The properties of such inhomogeneous metrics depend on the scalar field distribution in the extra dimensions.

  The extra-dimensional metrics represent a set of the cardinality of continuum even if the Largargian parameters are fixed. These metrics are stable under
  fluctuations in the extra space, as was shown in Section 4.

  It has been shown that the form of the stationary extra metric depends also on the value of the Hubble parameter $H$.
  The latter slowly changes with time in the early Universe. Therefore, we can approximate it as a constant and
  apply the obtained results under the assumption $H = \const$. As a result, the extra space metric and the
  scalar field distribution are changing during the inflationary period. This conclusion differs from a widespread
  point of view.

  Our analysis of the classical equations indicates that in the absence of matter fields only a maximally symmetric
  (spherical) metric in $\M_2$ is possible. This analytic result shows that the Ricci scalar of the extra space
  is unambiguously related to the Hubble parameter, and hence the extra-dimensional radius is slowly varying with
  time at the inflationary stage, and a similar picture might be expected for the present epoch.
  However, at present this radius has to be unacceptably large, this shortcoming being cured by invoking a scalar
  field, which makes its role very important.

  At high energy scales, quantum fluctuations perturb both the metric and the scalar field. It is widely assumed
  that our manifold was born at sub-Planckian energies, so that the scalar field randomly varies at those times.
  Let us estimate the conditions at which the fluctuations cannot disturb the extra space metric. Fluctuation
  are able to produce Kaluza-Klein excitations if the extra-dimensional scale $l_e$ is larger then the fluctuation
  wavelength $1/k$, where $k$ is magnitude of the its wave vector. For relativistic matter, the energy density
  $\rho\sim k^4$ while the Hubble parameter is $H\sim \sqrt{\rho/m_4^2}$. These estimates constrain the
  extra-dimensional scale as $l_e \lesssim 1/\sqrt{H m_4}$.

  This inequality allows us to impose a restriction on the extra space metric which is much stronger than those
  obtainable in collider experiments. Indeed, the cosmological constant
  and the gravitational constant do not vary within the present horizon. This means that fluctuations should be
  damped at the inflationary stage where $H=H_I\simeq 10^{13}$ GeV, so that the scale $l_e$ of the extra dimensions
  should be smaller than $1/\sqrt{H_Im_4}$. We conclude that the averaged size of the compact extra dimensions
  should be smaller than $\sim 10^{-28}$ cm. This limit confirms those obtained in \cite{Nikulin2019}, where
  it was shown that the slow roll motion of the inflaton is forbidden if $l_e > H^{-1}$.
  Recall that the collider limit is $l_e < 10^{-18}$ cm, which is 10 orders of magnitude weaker than our prediction.

  In conclusion, we would like to mention one more application of the idea of inhomogeneous
  extra dimensions. Consider, instead of \rf{ds2}, the 6D metric
\beq          \label{ds5}
		ds^2 = \e^{2\gamma(u)} \eta\mn dx^\mu dx^\nu  -  du^2 - r(u)^2 d\varphi^2,
\eeq
  where $\eta\mn$ is the 4D Minkowski metric, and the metric in $\M_2$ is the same as in \rf{ds2},
  but in terms of the metric \rf{ds1} we now have
\beq                  \label{b_A1}
             b_\mu = \gamma(u), \ \ \ \ b_4 = 0, \ \ \ \ b_5 = \ln r(u).
\eeq
  Using the expressions \rf{Ric_MM} and \rf{Ric_MN}, it is then straightforward to derive
  the explicit form of field equations for the unknowns $\gamma(u)$, $r(u)$ and $\phi(u)$.
  A tentative study has shown that there exist solutions with $u$-dependence of the circular
  radius $r$ somewhat similar to that shown in Fig.\,1 under boundary conditions
  similar to \rf{ini2}, \rf{ini3}, even if the scalar field is absent.
  If the size of $\M_2$ is large enough and with proper dependences $r(u)$
  and $\phi(u)$, the solutions can admit interpretations in terms of the brane world concept,
  somewhat similar to \cite{Bronnikov:2006bu, Bronnikov:2007kw}.
  Unlike solutions with the metric \rf{ds2}, mostly intended for the early Universe,
  those with \rf{ds5} are able to describe the present-day universe with very small 4D
  curvature, and a study of their possible properties and applications is in progress.

\section{Acknowledgments}

The work was supported by the Ministry of Science and Higher Education of the Russian Federation, Project "Fundamental properties of elementary particles and cosmology" N 0723-2020-0041.
The work of S.G.R. and  A.A.P. is performed according to the Russian Government Program of Competitive Growth of Kazan Federal University. The work of A.A.P was partly funded by the development program of the Regional Scientific and Educational Mathematical Center of the Volga Federal District, agreement N 075-02-2020.
The work of K.B. was partly funded by the RUDN University Program 5-100. The work of A.A.P. was also supported by the Russian Foundation for Basic Research Grant N 19-02-00496.
The work of S.G.R. is supported by the grant RFBR N~19-02-00930.


\begin{thebibliography}{10}
	
	\bibitem{Abbott:1984ba}
	R.~B. Abbott, S.~M. Barr and S.~D. Ellis, \emph{{Kaluza-Klein Cosmologies and
			Inflation}}, \href{https://doi.org/10.1103/PhysRevD.30.720}{\emph{Phys. Rev.}
		{\bfseries D30} (1984) 720}.
	
	\bibitem{Chaichian:2000az}
	M.~Chaichian and A.~B. Kobakhidze, \emph{{Mass hierarchy and localization of
			gravity in extra time}},
	\href{https://doi.org/10.1016/S0370-2693(00)00874-1}{\emph{Phys. Lett.}
		{\bfseries B488} (2000) 117}
	[\href{https://arxiv.org/abs/hep-th/0003269}{{\ttfamily hep-th/0003269}}].
	
	\bibitem{Randall:1999vf}
	L.~Randall and R.~Sundrum, \emph{{An Alternative to compactification}},
	\href{https://doi.org/10.1103/PhysRevLett.83.4690}{\emph{Phys. Rev. Lett.}
		{\bfseries 83} (1999) 4690}
	[\href{https://arxiv.org/abs/hep-th/9906064}{{\ttfamily hep-th/9906064}}].
	
	\bibitem{Brown:2013fba}
	A.~R. Brown, A.~Dahlen and A.~Masoumi, \emph{{Compactifying de Sitter space
			naturally selects a small cosmological constant}},
	\href{https://doi.org/10.1103/PhysRevD.90.124048}{\emph{Phys. Rev.}
		{\bfseries D90} (2014) 124048}
	[\href{https://arxiv.org/abs/1311.2586}{{\ttfamily 1311.2586}}].
	
	\bibitem{Fischbach:2009ud}
	E.~Fischbach, G.~L. Klimchitskaya, D.~E. Krause and V.~M. Mostepanenko,
	\emph{{On the validity of constraints on light elementary particles and
			extra-dimensional physics from the Casimir effect}},
	\href{https://doi.org/10.1140/epjc/s10052-010-1326-2}{\emph{Eur. Phys. J.}
		{\bfseries C68} (2010) 223}
	[\href{https://arxiv.org/abs/0911.1950}{{\ttfamily 0911.1950}}].
	
	\bibitem{2018GrCo...24..154B}
	S.~V. {Bolokhov} and K.~A. {Bronnikov}, \emph{{On cosmology in nonlinear
			multidimensional gravity with multiple factor spaces}},
	\href{https://doi.org/10.1134/S0202289318020044}{\emph{Grav. Cosmol.}
		{\bfseries 24} (2018) 154}
	[\href{https://arxiv.org/abs/1803.04904}{{\ttfamily 1803.04904}}].
	
	\bibitem{1994CQGra..11.2483V}
	H.~{van Elst}, J.~E. {Lidsey} and R.~{Tavakol}, \emph{{Quantum cosmology and
			higher-order Lagrangian theories}},
	\href{https://doi.org/10.1088/0264-9381/11/10/008}{\emph{Classical Quant.
			Grav.} {\bfseries 11} (1994) 2483}
	[\href{https://arxiv.org/abs/arXiv:gr-qc/9404044}{{\ttfamily
			arXiv:gr-qc/9404044}}].
	
	\bibitem{Bousso:1998ed}
	R.~Bousso and A.~D. Linde, \emph{{Quantum creation of a universe with $\Omega
			\neq 1$: Singular and nonsingular instantons}},
	\href{https://doi.org/10.1103/PhysRevD.58.083503}{\emph{Phys. Rev.}
		{\bfseries D58} (1998) 083503}
	[\href{https://arxiv.org/abs/gr-qc/9803068}{{\ttfamily gr-qc/9803068}}].
	
	\bibitem{2009arXiv0909.2566H}
	J.~J. {Halliwell}, \emph{{Introductory Lectures on Quantum Cosmology (1990)}},
	{\emph{ArXiv e-prints} (2009) }
	[\href{https://arxiv.org/abs/0909.2566}{{\ttfamily 0909.2566}}].
	
	\bibitem{yurov2005quantum,
    title={Quantum Creation of a Universe in Albrecht-Magueijo-Barrow model},
    author={A. V. Yurov and V. A. Yurov},
    year={2005},
    eprint={hep-th/0505034},
    archivePrefix={arXiv},
    primaryClass={hep-th}
}

	
	\bibitem{Brandenberger:2006vv}
	R.~H. Brandenberger, A.~Nayeri, S.~P. Patil and C.~Vafa, \emph{{String gas
			cosmology and structure formation}},
	\href{https://doi.org/10.1142/S0217751X07037159}{\emph{Int. J. Mod. Phys.}
		{\bfseries A22} (2007) 3621}
	[\href{https://arxiv.org/abs/hep-th/0608121}{{\ttfamily hep-th/0608121}}].
	
	\bibitem{Tegmark:2005dy}
	M.~Tegmark, A.~Aguirre, M.~Rees and F.~Wilczek, \emph{{Dimensionless constants,
			cosmology and other dark matters}},
	\href{https://doi.org/10.1103/PhysRevD.73.023505}{\emph{Phys. Rev.}
		{\bfseries D73} (2006) 023505}
	[\href{https://arxiv.org/abs/astro-ph/0511774}{{\ttfamily
			astro-ph/0511774}}].
	
	\bibitem{Gani:2014lka}
	V.~A. Gani, A.~E. Dmitriev and S.~G. Rubin, \emph{{Deformed compact extra space
			as dark matter candidate}},
	\href{https://doi.org/10.1142/S0218271815450017}{\emph{Int. J. Mod. Phys.}
		{\bfseries D24} (2015) 1545001}
	[\href{https://arxiv.org/abs/1411.4828}{{\ttfamily 1411.4828}}].
	
	\bibitem{Rubin:2016ude}
	S.~G. Rubin, \emph{{Inhomogeneous extra space as a tool for the top-down
			approach}}, \href{https://doi.org/10.1155/2018/2767410}{\emph{Adv. High
			Energy Phys.} {\bfseries 2018} (2018) 2767410}
	[\href{https://arxiv.org/abs/1609.07361}{{\ttfamily 1609.07361}}].
	
	\bibitem{Rubin:2015pqa}
	S.~G. Rubin, \emph{{Scalar field localization on deformed extra space}},
	\href{https://doi.org/10.1140/epjc/s10052-015-3553-z}{\emph{Eur. Phys. J.}
		{\bfseries C75} (2015) 333}
	[\href{https://arxiv.org/abs/1503.05011}{{\ttfamily 1503.05011}}].
	
	\bibitem{Lyakhova:2018zsr}
	Y.~Lyakhova, A.~A. Popov and S.~G. Rubin, \emph{{Classical evolution of
			subspaces}}, \href{https://doi.org/10.1140/epjc/s10052-018-6251-9}{\emph{Eur.
			Phys. J.} {\bfseries C78} (2018) 764}
	[\href{https://arxiv.org/abs/1807.06235}{{\ttfamily 1807.06235}}].
	
	\bibitem{Nikulin2019}
	V.~Nikulin and S.~G. Rubin, \emph{Inflationary limits on the size of compact
		extra space},
	\href{https://doi.org/10.1142/S0218271819410049}{\emph{International Journal
			of Modern Physics D} {\bfseries 28} (2019) 1941004}
	[\href{https://arxiv.org/abs/1903.05725}{{\ttfamily 1903.05725}}].
	
	\bibitem{Starobinsky:1980te}
	A.~A. Starobinsky, \emph{{A New Type of Isotropic Cosmological Models Without
			Singularity}},
	\href{https://doi.org/10.1016/0370-2693(80)90670-X}{\emph{Phys. Lett.}
		{\bfseries B91} (1980) 99}.
	
	\bibitem{DeFelice:2010aj}
	A.~De~Felice and S.~Tsujikawa, \emph{{f(R) theories}},
	\href{https://doi.org/10.12942/lrr-2010-3}{\emph{Living Rev. Rel.} {\bfseries
			13} (2010) 3} [\href{https://arxiv.org/abs/1002.4928}{{\ttfamily
			1002.4928}}].
	
	\bibitem{2014JCAP...01..008B}
	K.~{Bamba}, A.~N. {Makarenko}, A.~N. {Myagky}, S.~{Nojiri} and S.~D.
	{Odintsov}, \emph{{Bounce cosmology from F(R) gravity and F(R) bigravity}},
	\href{https://doi.org/10.1088/1475-7516/2014/01/008}{\emph{J. Cosmol.
			Astropart. Phys.} {\bfseries 1} (2014) 8}
	[\href{https://arxiv.org/abs/1309.3748}{{\ttfamily 1309.3748}}].
	
	\bibitem{Nojiri_2017}
	S.~Nojiri, S.~Odintsov and V.~Oikonomou, \emph{Modified gravity theories on a
		nutshell: Inflation, bounce and late-time evolution},
	\href{https://doi.org/10.1016/j.physrep.2017.06.001}{\emph{Physics Reports}
		{\bfseries 692} (2017) 1–104}.
	
	\bibitem{2007PhLB..651..224N}
	S.~{Nojiri}, S.~D. {Odintsov} and P.~V. {Tretyakov}, \emph{{Dark energy from
			modified F(R)-scalar-Gauss Bonnet gravity}},
	\href{https://doi.org/10.1016/j.physletb.2007.06.029}{\emph{Physics Letters
			B} {\bfseries 651} (2007) 224}
	[\href{https://arxiv.org/abs/0704.2520}{{\ttfamily 0704.2520}}].
	
	\bibitem{Bronnikov:2005iz}
	K.~A. Bronnikov and S.~G. Rubin, \emph{{Self-stabilization of extra
			dimensions}}, \href{https://doi.org/10.1103/PhysRevD.73.124019}{\emph{Phys.
			Rev.} {\bfseries D73} (2006) 124019}
	[\href{https://arxiv.org/abs/gr-qc/0510107}{{\ttfamily gr-qc/0510107}}].
	
	\bibitem{BronRub}
	K.~A. {Bronnikov} and S.~G. {Rubin}, \emph{Black Holes, Cosmology and Extra
			Dimensions}, World Scientific Publishing Co. Pte. Ltd., 2013.
	
	\bibitem{2003Ap&SS.283..679G}
	U.~{G{\"u}nther}, P.~{Moniz} and A.~{Zhuk}, \emph{{Multidimensional Cosmology
			and Asymptotical AdS}},
	\href{https://doi.org/10.1023/A:1022532313230}{\emph{Astrophys. Space Sci.}
		{\bfseries 283} (2003) 679}
	[\href{https://arxiv.org/abs/arXiv:gr-qc/0209045}{{\ttfamily
			arXiv:gr-qc/0209045}}].
	
	\bibitem{Bronnikov:2008cr}
	K.~Bronnikov, S.~Kononogov, V.~Melnikov and S.~Rubin, \emph{{Cosmologies from
			nonlinear multidimensional gravity with acceleration and slowly varying G}},
	\href{https://doi.org/10.1134/S0202289308030043}{\emph{Grav.\ Cosmol.}
		{\bfseries 14} (2008) 230} [\href{https://arxiv.org/abs/0804.0973}{{\ttfamily
			0804.0973}}].
	
	\bibitem{Bronnikov:2009zza}
	K.~Bronnikov, S.~Rubin and I.~Svadkovsky, \emph{{High-order multidimensional
			gravity and inflation}},
	\href{https://doi.org/10.1134/S0202289309010083}{\emph{Grav.\ Cosmol.}
		{\bfseries 15} (2009) 32}.
	
	\bibitem{Bronnikov:2009ai}
	K.~Bronnikov, S.~Rubin and I.~Svadkovsky, \emph{{Multidimensional world,
			inflation and modern acceleration}},
	\href{https://doi.org/10.1103/PhysRevD.81.084010}{\emph{Phys.\ Rev.\ D}
		{\bfseries 81} (2010) 084010}
	[\href{https://arxiv.org/abs/0912.4862}{{\ttfamily 0912.4862}}].
	
	\bibitem{2018EPJC...78..373P}
	S.~A. {Pavluchenko} and A.~{Toporensky}, \emph{{Effects of spatial curvature
			and anisotropy on the asymptotic regimes in Einstein-Gauss-Bonnet gravity}},
	\href{https://doi.org/10.1140/epjc/s10052-018-5855-4}{\emph{European Physical
			Journal C} {\bfseries 78} (2018) 373}
	[\href{https://arxiv.org/abs/1709.04258}{{\ttfamily 1709.04258}}].
	
	\bibitem{2010IJGMM..07..797I}
	V.~D. {Ivashchuk}, \emph{{On cosmological-type solutions in multidimensional
			model with Gauss-Bonnet term}},
	\href{https://doi.org/10.1142/S0219887810004555}{\emph{International Journal
			of Geometric Methods in Modern Physics} {\bfseries 7} (2010) 797}
	[\href{https://arxiv.org/abs/0910.3426}{{\ttfamily 0910.3426}}].
	
	\bibitem{Bronnikov:2006bu}
	K.~Bronnikov, B.~Meierovich and S.~Abdyrakhmanov, \emph{{Global topological
			defects in extra dimensions and the brane world concept}}, {\emph{Grav.\
			Cosmol.} {\bfseries 12} (2006) 106}.
	
	\bibitem{Bronnikov:2007kw}
	K.~Bronnikov and B.~Meierovich, \emph{{Global strings in extra dimensions: A
			full map of solutions, matter trapping and the hierarchy problem}},
	\href{https://doi.org/10.1007/s11447-008-2005-0}{\emph{J.\ Exp.\ Theor.\
			Phys.} {\bfseries 106} (2008) 247}
	[\href{https://arxiv.org/abs/0708.3439}{{\ttfamily 0708.3439}}].
	
\end{thebibliography}


\providecommand{\href}[2]{#2}\begingroup\raggedright\endgroup

\end{document}